\documentclass[aps,prl,groupedaddress]{revtex4-1}
\usepackage{graphics}

\usepackage{epsfig}
\usepackage{amsmath}
\usepackage{booktabs}
\usepackage{array}
\usepackage{amssymb}
\usepackage{amsfonts}
\usepackage{dsfont}
\usepackage{caption}
\usepackage[debug]{mciteplus}
\usepackage{braket}
\usepackage{color}
\usepackage{ulem}
\usepackage{bm}
\newcommand{\1}{\mbox{1}\hspace{-0.25em}\mbox{l}}
\newcommand{\Xbar}{\overline{X}}
\newcommand{\Ybar}{\overline{Y}}
\def\Ketbra#1{\Ket{#1}\Bra{#1}}
\begin{document}

\title{Gate fidelity and coherence of an electron spin in a Si/SiGe quantum dot with micromagnet}

\hsize\textwidth\columnwidth\hsize\csname@twocolumnfalse\endcsname

\author{E. Kawakami$^1$, T. Jullien$^1$, P. Scarlino$^1$, D. R. Ward$^2$, D. E. Savage$^2$, M. G. Lagally$^2$, V. V. Dobrovitski$^3$, Mark Friesen$^2$, S. N. Coppersmith$^2$, M. A. Eriksson$^2$, and L. M. K. Vandersypen$^{1,4}$}
\affiliation{$^1$QuTech and Kavli Institute of Nanoscience, TU Delft, Lorentzweg 1, 2628 CJ Delft, The Netherlands\\
$^2$University of Wisconsin-Madison, Madison, WI 53706, USA\\
$^3$Ames Laboratory, U.S. DOE, Iowa State University, Ames Iowa 50011, USA \\
$^4$Components Research, Intel Corporation, 2501 NW 29th Ave, Hillsboro, OR 97124, USA
}

%
%
%
%

\begin{abstract}
The gate fidelity and the coherence time of a qubit are important benchmarks for quantum computation. We construct a qubit using a single electron spin in a Si/SiGe quantum dot and control it electrically via an artificial spin-orbit field from a micromagnet. We measure an average single-qubit gate fidelity of $\approx$ 99$\%$ using randomized benchmarking, which is consistent with dephasing from the slowly evolving nuclear spins in substrate. The coherence time measured using dynamical decoupling extends up to $\approx$ 400 $\mu$s  for 128 decoupling pulses, with no sign of saturation. We find evidence that the coherence time is limited by noise in the 10 kHz $-$ 1 MHz range, possibly because charge noise affecting the spin via the micromagnet gradient. This work shows that an electron spin in a Si/SiGe quantum dot is a good candidate for quantum information processing as well as for a quantum memory, even without isotopic purification. 
\end{abstract}
\maketitle

\section{Introduction}
The performance of a qubit is characterized by how accurately operations on the qubit are implemented and for how long its state is preserved. For improving qubit performance, it is important to identify the nature of the noise which introduces gate errors and leads to loss of qubit coherence. Ultimately, what counts is to balance the ability to drive fast qubit operations and the need for long coherence times \cite{Fowler2012}.

Electron spins in Si quantum dots are now known to be one of the most promising qubit realizations for their potential to scale up and their long coherence times \cite{Zwanenburg2013,Veldhorst2014, Veldhorst2015,Muhonen2014,Kim2014,Kawakami2014,Maune2012,Tyryshkin2012,Morton2011}. Using magnetic resonance on an electron spin bound to a phosphorous impurity in isotopically purified $^\mathrm{28}$Si \cite{Muhonen2014} or confined in a $^\mathrm{28}$Si MOS quantum dot \cite{Veldhorst2014} $\approx 0.3$ MHz Rabi frequencies, gate fidelities over 99.5\%, and spin memory times of tens to hundreds of ms have been achieved. Also electrical control of an electron spin has been demonstrated in a (natural abundance) Si/SiGe quantum dot. This was achieved by applying an AC electric field that oscillates the electron wave function back and forth in the gradient magnetic field of a local micromagnet \cite{Kawakami2014}. The advantage of electrical control over magnetic control is that electric fields can be generated without the need for microwave cavities or striplines and allows better spatial selectivity, which simplifies individual addressing of qubits. However, the magnetic field gradient also makes the qubit sensitive to electrical noise, so it is important to examine whether the field gradient limits the spin coherence time and the gate fidelity. 

In our previous work \cite{Kawakami2014}, the effect of electrical noise on spin coherence and gate fidelity was overwhelmed by transitions between the lowest two valley-orbit states. Since different valley-orbit states have slightly different Larmor frequencies, such a transition will quickly randomize the phase of the electron spin. If valley-orbit transitions can be (largely) avoided, then the question becomes what limits coherence and fidelities instead.

Here we measure the gate fidelity and spin echo times for an electron spin in a Si/SiGe quantum dot in a regime where the electron stably remains in the lowest valley-orbit state for long times, and where the corresponding resonance condition is well separated from that associated with the other valley-orbit state. In order to learn more about the dominant noise sources in this new regime, we use dynamical decoupling experiments to extract the noise spectrum in the range of 5 kHz -1 MHz, and we compare this spectrum with spectra derived from numerical simulations for various noise sources. We also study the influence of the various noise sources on the gate fidelity.

\section{Device and Measurement Setup}
 A single electron spin is confined in a gate-defined quantum dot in an undoped Si/SiGe heterostructure \cite{Kim2014,Kawakami2014,Maune2012} (Fig.~1). The sample is attached to the mixing chamber (MC) stage of a dilution refrigerator with base temperature of $\approx$ 25 mK, and subject to a static external magnetic field of 794.4 mT along the direction as indicated in the inset of Fig.~1.  Spin rotations are achieved by applying microwave excitation to one of the gates, which oscillates the electron wave function back and forth in the magnetic field gradient produced by two cobalt micromagnets fabricated on top of the device. The device used in this work is the same as in the previous work \cite{Kawakami2014} but the applied gate voltages are set differently to obtain a higher valley-orbit splitting.

\begin{figure}[!h]
\includegraphics[width=11cm] {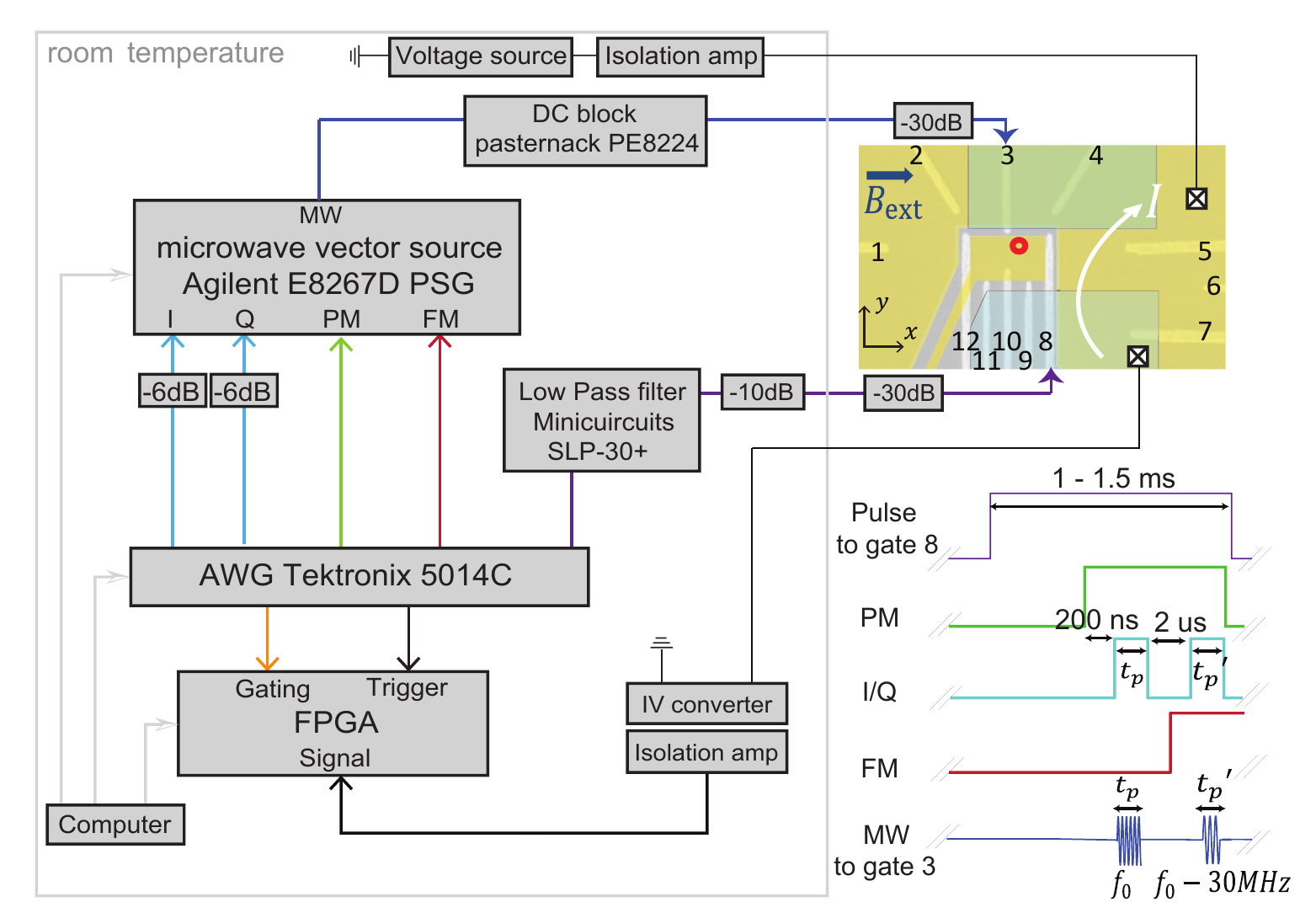} 
\caption{\label{fig:fig1} Device schematic and block diagram depicting the generation of gate voltage pulses and phase-controlled microwave bursts, and read-out trace analysis and post analysis using an FPGA. The main components are described in the text. For the I/Q inputs, 6 dB attenuators are added to reduce the noise from the AWG. In order to reduce the noise going to the sample from the AWG, a Minicircuit low pass filter SLP-30+ and a 10 dB attenuator are added at room temperature. A Pasternack DC block PE8224 is added at room temperature behind the microwave source to reduce  low frequency noise. The 30 dB attenuation at low temperature is divided over a 20 dB attenuator at the 1 K plate and a 10 dB attenuator at the MC stage for each of two high-frequency lines, connected to gate 3 and gate 8. One of the two ohmic contacts of the sensing dot is connected to a room-temperature voltage source and the other is connected to the input of a homemade JFET current-to-voltage (IV) converter via RC and copper powder filters mounted at the MC stage and pi-filters at room temperature (not shown in the figure). The output voltage signal of the IV converter is digitized and processed by an FPGA. A gating pulse sent to the FPGA defines the segment of the signal that is to be analyzed. An additional trigger pulse is applied to the FPGA before the entire sequence starts. The inset shows the voltage pulse applied to gate 8 (purple line), the pulses used for PM (green line), gating the FPGA (orange line), I/Q modulation (light blue line), and FM (red line), and the microwave burst applied to gate 3 (blue line) during the manipulation stage.}
\end{figure} 

 The measurement scheme consists of 4 stages: initialization, manipulation, read-out and emptying, as shown in \cite{Kawakami2014}. Differently from \cite{Kawakami2014}, the 4-stage voltage pulse is applied to gate 8 and the microwave excitation is applied to gate 3. The initialization and read-out stages take 4-5 ms and the manipulation and emptying stages last 1-1.5 ms. 
 
 Since the experimental details of the set-up are important for the results below, we here summarize the key components. A voltage pulse applied to the gate 8 is generated by an arbitrary waveform generator (Tektronix AWG 5014C). Phase-controlled microwave bursts are generated by an Agilent microwave vector source E8267D with the I (in-phase) and Q (out-of-phase) components controlled by two channels of the AWG. The on/off ratio of the I/Q modulation is 40 dB. If the microwave power arriving at the sample is not sufficiently suppressed in the “off” state, the control fidelity is reduced and the effective electron temperature increases, which in turn will result in lower read-out and initialization fidelities. Reduced fidelities were indeed observed when applying high power microwave excitation ($>15$ dBm at the source) using I/Q modulation only. As a solution, we use digital pulse modulation (PM) in series with the I/Q modulation, which gives a total on/off ratio of $\approx$ 120 dB. A drawback of PM is that the switching rate is lower. Therefore, the PM is turned on 200 ns before the I/Q modulation is turned on  (see inset of Fig.~1). We also observe that the total microwave burst time applied to the sample per cycle affects the read-out and initialization fidelities (SI Appendix). In order to keep the read-out and initialization fidelities constant we apply an off-resonance microwave burst (with microwave frequency detuned by 30 MHz from the resonance frequency) 2 $\mu$s after the on-resonance microwave burst, so that the combined duration of the two bursts is fixed. To achieve this rapid shift of the microwave frequency, we used Frequency Modulation (FM) controlled by another channel of the AWG. FM is turned on 1 $\mu$s after the on-resonance burst is turned off (see inset of Fig.~1). 
 
The electron spin state is read out via spin-to-charge conversion combined with real-time charge detection \cite{Elzerman2004}. The probability that the current exceeds a predefined threshold during the read-out stage is interpreted as the spin-up probability of the electron \cite{Kawakami2014}. The analysis of the real-time traces and the statistical analysis of the read-out events are done on-the-fly using a field-programmable gate array (FPGA) as depicted in Fig.~1. This allows us to measure faster without waiting for the transfer of real-time traces to a computer. Data points were taken by cycling through the various burst times, spin echo waiting times, or randomized gate sequences, and repeating these entire cycles 250 - 1000 times. This order of the measurements helps to suppress artifacts in the data caused by slow drift in the set-up or sample.

\section{High-quality Rabi Oscillations}

Rabi oscillations are recorded by varying the burst time and the microwave frequency. With the present gate voltage settings, the spin resonance frequencies corresponding to the two lowest valley-orbit states are separated by $\approx$ 5 MHz (at $B_{\mathrm{ext}}=794.4$ mT), so that two well separated chevron patterns characteristic for Rabi oscillations are observed [see Fig.~2(a)]. This difference of $\approx$ 5 MHz results mainly from slightly different electron $g$-factors between the two valley-orbit states. The population of the valley-orbit ground state is estimated to be $\approx 80\%$ from Fig.~2(a), which is higher than in our previous work \cite{Kawakami2014}, and implies a higher valley-orbit splitting. Fig.~2(b) shows a Rabi oscillation of a single spin with the electron in the ground valley-orbit state. The Rabi frequency extracted from the data is $1.345$ MHz. The decay of the oscillation is what we would expect assuming a statistical distribution of resonance conditions with a line width of 0.63 MHz (FWHM), which is the number extracted from the continuous wave response (not shown). This line width corresponds to $T_2^* \approx 1$ $\mu$s, and is presumably dominated by the 4.7\% $^\mathrm{29}$Si spins in the substrate, similar to \cite{Kawakami2014}. Here there is no evidence of additional decay mechanisms. In particular, we do not see any indication of intervalley switching or the combined effects of electrical noise and the magnetic field gradient.

\begin{figure}[!h]
\includegraphics[width=11cm] {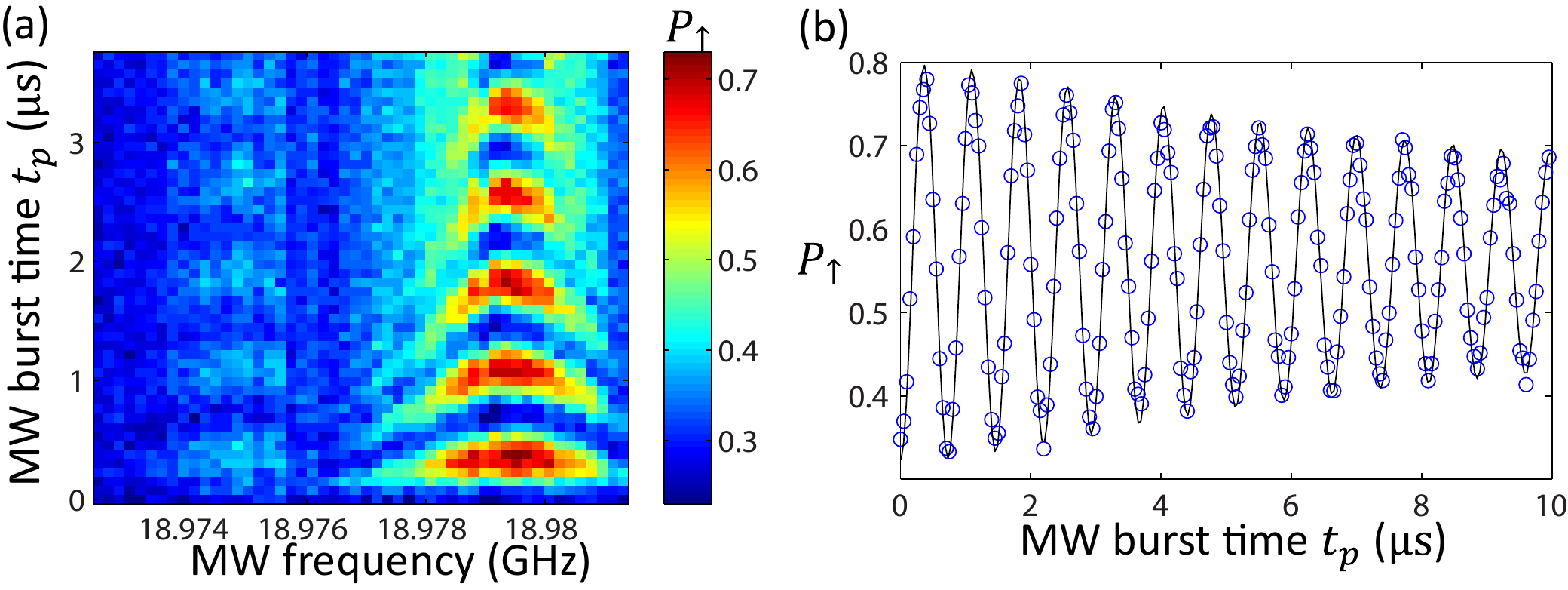}
\caption{\label{fig:fig2} (a) Measured spin-up probability, $P_{\uparrow}$, as a function of $f_{MW}$ and burst time $t_p$ (microwave power at the source $P=18.85$ dBm), showing two Rabi chevron patterns corresponding to the two valley-orbit states. The resonance frequency of the valley-orbit ground state is 18.9795 GHz and that of the excited state is 18.9750 GHz. The signal coming from the excited state is much smaller due to its lower population. (b) Measured spin-up probability, $P_{\uparrow}$, showing a Rabi oscillation for the ground valley-orbit state (blue circles). During the manipulation stage, on-resonance microwave excitation (at $f_{MW}=18.9795$ GHz) was applied for a time $t_p$ and off-resonance microwave ($f_{MW}=18.9195$ GHz) was applied for a time $t_p'=10\mu$s $-t_p$, in order to keep the total duration of the microwave bursts fixed to 10 $\mu$s for every data point. The black line shows a numerical fit with a model that includes  a constant driving field in the rotating frame (which is a fit parameter) and (quasi-)static noise modeled by a Gaussian distribution of resonance offsets with width 0.63 MHz (FWHM).}
\end{figure} 

\section{Dynamical Decoupling}

Next we examine the spin memory time of this electrically controlled spin qubit. In our previous work \cite{Kawakami2014,Scarlino2014}, due to switching between the two valley-orbit states, the Hahn echo decay was exponential with coherence time $\approx 40$ $\mu$s. Furthermore, we were unable to extend the coherence time using multiple echo pulses. Due to the difference in Larmor frequency between two valley-orbit states, as soon as a switch from one to the other valley-orbit state occured, phase information could not be recovered by echo pulses. In this work, we observe significantly extended coherence times presumably because the switching between valleys is slower in the present gate voltage configuration. 

We study the spin memory characteristics using two types of two-axis dynamical decoupling sequences, based on the XY4 \cite{Maudsley1986}, (XY4)$^n$ (sometimes called vCDD \cite{Alvarez2012}) and XY8 \cite{Gullion1990} protocols. The insets in Fig.~3(a,b) show the (XY4)$^n$ and XY8 pulse sequences for 16 $\pi$ pulses. We use $X$ and $Y$ to denote $\pi$ rotations about $\hat{x}$ and $\hat{y}$, and $\bar{X}$ and $\bar{Y}$ for rotations about $-\hat{x}$ and $-\hat{y}$. Such two-axis decoupling sequences are chosen in order to reduce the effect of pulse imperfections and to equally preserve the spin components along all directions in the $\hat{x}-\hat{y}$ plane  \cite{DeLange2010}, which is important for quantum information processing. One-axis decoupling sequences such as CPMG \cite{Carr1954,Meiboom1958} may artificially preserve a specific spin component for a longer time than two-axis decoupling sequences, but with a reduced coherence time of the orthogonal spin component \cite{Souza2012,Wang2012,Bluhm2010}. The visibility of the echo amplitude decreases for larger numbers of $\pi$ pulses, $N_\pi$, due to the pulse imperfections. Therefore, to facilitate direct comparison of the decay rates with different numbers of $\pi$ pulses, in Fig.~3(a,b) we show the data, normalized to the echo amplitude at $t_\mathrm{wait}=0$, as a function of the total wait time $t_\mathrm{wait}$ for (XY4)$^n$ and XY8, respectively. 

To analyze these decay curves, we adopt a semiclassical approach, in which the decay curve of the echo amplitude is written as
\begin{equation}
P(t_\mathrm{wait})=\exp\left[-W(t_\mathrm{wait})\right] \label{Decay} 
\end{equation}
with
\begin{equation}
W(t_\mathrm{wait})= \int_{-\infty}^{\infty} \frac{S(\omega)}{2\pi} \frac{F(\omega)}{\omega^2} d\omega. \label{W} 
\end{equation}
$S(\omega)$ is the noise spectrum that produces an effective magnetic field fluctuation $\delta b(t)$ along the same direction as the quantization axis \cite{deSousa2009}. More concretely, the relation between $S(\omega)$ and $\delta b(t)$ is described as $S(\omega) = \int_{-\infty}^{\infty} \gamma_e^2 \langle \delta b(0) \delta b(t) \rangle  e^{\mathrm{i}\omega t} dt$ with $\gamma_e$ the gyro-magnetic ratio of the electron. $F(\omega)$ is the filter function of the pulse sequence \cite{Uhrig2007,Cywinski2008}. First we assume that the noise spectrum dominating the decoherence is described by a power law,
\begin{eqnarray}
S(\omega)=\frac{K}{\omega^{\alpha-1}}, \label{singlepowerSomega}
\end{eqnarray}
as seen in GaAs quantum dots \cite{Medford2012} and NV centers in diamond \cite{DeLange2010}. Under this assumption, if the filter function $F(\omega)$ is sufficiently narrow around $\omega=\frac{\pi N_\pi}{t_{\mathrm{wait}}} $ (which we verified is the case for $N_\pi \geq 4$), the decay curve can be written as \cite{Bylander2011}
\begin{eqnarray}
P(t_\mathrm{wait})=\exp\left[-\left(\frac{t_\mathrm{wait}}{T_2}\right)^\alpha\right], \label{DecayAlpha} 
\end{eqnarray}
with $T_2=T_2^0 N_\pi ^{(1-\frac{1}{\alpha})}$ and $T_2^0=\left( \frac{2}{K}\right)^\frac{1}{\alpha} \pi^{1-\frac{1}{\alpha}}$. Fig.~3(c) shows $T_2$ as a function of the number of $\pi$ pulses obtained by fitting Eq. [4] to the decay curves. The longest $T_2$ time reached is $\approx 400 \mu$s with XY8 and $N_\pi=128$ (data shown in SI Appendix). We fitted $T_2=T_2^0 N_\pi ^{(1-\frac{1}{\alpha})}$ to the data (leaving out the case $N_\pi = 1$, the Hahn echo) and the resulting fit is shown in green in Fig.~3(c). 

We can derive the noise spectrum from the decay curves in Fig.~3(a,b) using the fact that the filter function is narrow around $\omega=\frac{\pi N_\pi}{t_{\mathrm{wait}}} $ for $N_\pi \geq 4$, \cite{deSousa2009} (SI Appendix). The circles in Fig.~3(d) show the noise spectrum extracted from six decay curves in Fig.~3(b). The colors of the circles in Fig.~3(d) correspond to the colors used in Fig.~3(b) for different $N_\pi$. The green solid line in Fig.~3(d) is based on Eq.~\eqref{singlepowerSomega} with $T_2^0$ and $\alpha$ obtained from the fit (green line) to the data in Fig.~3(c). Its decay is close to a $1/f$ decay. Although this line shows an overall good agreement with the noise spectrum extracted from the decay curves, it does not match with the flat region at $\omega/2\pi \lesssim 30$ kHz. 

In order to capture both the flat and decaying parts of the spectrum and obtain more insight into the nature of the noise spectrum, we now write the noise spectrum in the form 
\begin{eqnarray}
S(\omega)=\frac{A}{1+(\omega \tau_c)^{\alpha-1}} \label{OU_1overf}. 
\end{eqnarray}
We fit Eq.~\eqref{Decay} to the six decay curves (leaving out $N_\pi=1$) in Fig.~3(b) simultaneously, using also Eq.~\eqref{W} and Eq.~\eqref{OU_1overf}, with $A$ and $\tau_c$ as the only fitting parameters. We first perform this fit (numerically) using $\alpha=2$, close to the previously fitted value $\alpha=1.8$ obtained using Eq.~\eqref{singlepowerSomega}, but the fits deviate from the measured echo decays (see Fig.~S5(e)). A better fit to the echo decay data using Eq.~\eqref{OU_1overf} is obtained for $\alpha=3$ (Fig.~3(b)), in which case Eq.~\eqref{W} can be expressed analytically \cite{Wang2012}. The fits in Fig.~3(b) yield $A=(2.5 \pm 0.2) \cdot 10^4$ $\mathrm{rad}^2$s$^{-1}$ and $\tau_c=2.46 \pm 0.17$ $\mu$s. The resulting fit, plotted as a thick black line in Fig.~3(d), shows reasonable agreement with $S(\omega)$ obtained from the experimental data. 

Extrapolating the fitted noise spectrum to frequencies below 5 kHz, where we do not have experimental data (dashed line in Fig.~3(d)), the noise spectral density looks flat; this would result in an exponential Ramsey decay with $T_2^*\approx 80$ $\mu$s \cite{deSousa2009}. However, the measured Ramsey decay is Gaussian and has a much shorter $T_2^* \approx 1 \mu$s. Therefore the noise power at low frequencies must exceed the dashed horizontal black line in Fig.~3(d) (SI Appendix).

\begin{figure}[!h]
\includegraphics[width=11cm] {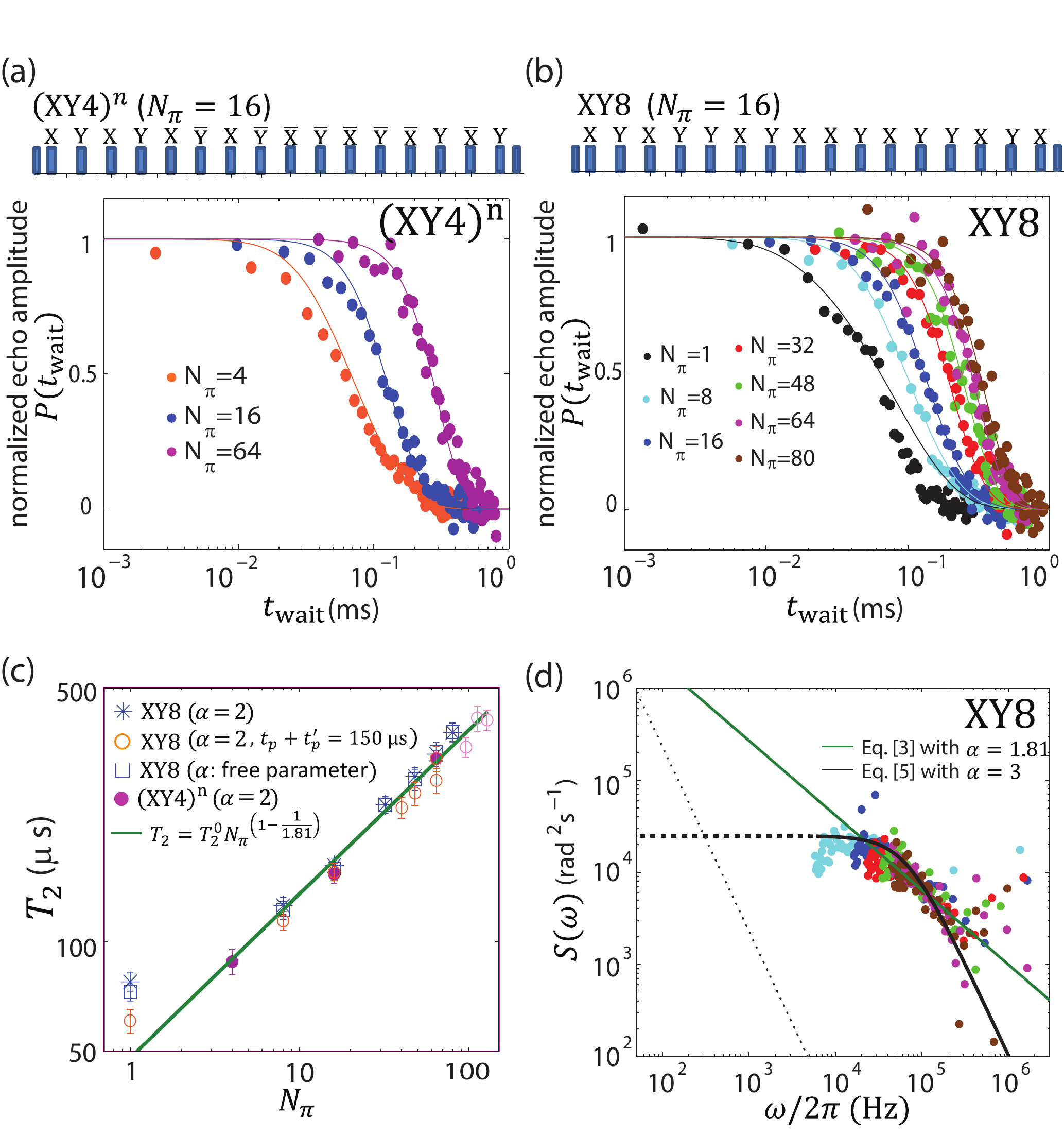}
\caption{\label{fig:fig3}  (a) Normalized spin echo amplitude as a function of the total waiting time $t_\mathrm
{wait}$ using the (XY4)$^n$ pulse sequence for $N_{\pi}=4$ (orange points), $16$ (blue points), and $64$ (purple points) pulses (concatenated level $n=$1, 2 and 3, respectively). The inset shows the (XY4)$^n$ pulse sequence for $N_{\pi}=16$ ($n=2$). The first and the last pulses are $\pi/2$ pulses and the 16 pulses in the middle are $\pi$ pulses. X, $\bar{\mathrm{X}}$, $\bar{\mathrm{Y}}$ or Y indicate the phase of the pulses. The solid lines present fits using Eq.~\eqref{Decay} with Eq.~\eqref{W} and Eq.~\eqref{OU_1overf} for $\alpha=3$. 
(b) Similar to (a) but using a Hahn echo sequence for $N_{\pi}=1$ and a XY8 sequence for $N_{\pi}=8,16,32,48,64,80$. The inset shows XY8 pulse sequence for $N_{\pi}=16$. Fits as in (a), except that the solid line for $N_\pi=1$ is the decay curve with $A$ and $\tau_c$ obtained from the fit to the other 6 decay curves in (b). 
(c) Coherence time, $T_2$, as a function of the number of $\pi$ pulses $N_{\pi}$ using XY8 (blue asterisks, blue squares and orange circles) and (XY4)$^n$ (purple circles). See the SI Appendix for the pulse sequences used for pink open  circles. The $T_2$ values are obtained by fitting Eq.~\eqref{DecayAlpha} to the decay curves. The choice of $\alpha$ did not much affect the extracted $T_2$. The values shown are for $\alpha=2$ except for the blue squares for which $\alpha$ is left as a fitting parameter. The green line presents a fit to the data (leaving out $N_{\pi}=1$) using $T_2=T_2^0 N_\pi^{(1-1/\alpha)}$. From this fit, we obtained $T_2^0=48\pm 8$ $\mu$s and $\alpha=1.81 \pm 0.14$. 
(d) Noise spectrum extracted from Fig.~3(b). The green solid line corresponds to Eq.~\eqref{singlepowerSomega} with $T_2^0=48$ $\mu$s and $\alpha=1.81$. The black line presents a fit using Eq.~\eqref{OU_1overf}, see main text. The dotted black line represents the calculated noise spectrum produced by the $^{29}$Si nuclear spin dynamics (see SI Appendix for the details of the calculation).}
\end{figure} 

We now turn to the noise mechanisms and examine whether the hyperfine coupling of the electron spin with the evolving nuclear spins can explain the observed noise spectrum. Nuclear spin dynamics has two main mechanisms, hyperfine-mediated and dipole-dipole interactions between nuclear spins. Decoherence due to the hyperfine-mediated interactions is negligible in Si at $B\approx$ 800 mT \cite{Cywinski2009} (SI Appendix). However, magnetic dipole-dipole induced nuclear spin dynamics cannot be neglected. We performed numerical simulations of  the spectrum of the nuclear spin noise and of the Hahn echo decay for a dot with 4.7\% of $^\mathrm{29}$Si nuclei (natural abundance) within the coupled pair-cluster expansion \cite{Yang2009} for several choices of the quantum dot parameters (see SI Appendix for details). A representative calculated spectrum is shown by the dotted line in Fig.~3(d). The measured Gaussian-shape Ramsey decay with $T_2^* \sim 1 ~\mu$s is consistent with this spectrum so presumably the randomly oriented $^\mathrm{29}$Si nuclear spins dominate the noise at low frequencies \cite{Kawakami2014,deSousa2009}. They also dominate the gate fidelities discussed below. However, at higher frequencies the noise spectrum calculated from the nuclear spin dynamics is far below the measured noise spectrum and we consistently found a calculated Hahn echo decay time $T_2$ above 0.5 ms, which is much longer than the measured value of 70 $\mu$s. Nuclear spin noise thus cannot explain the observed Hahn echo decay.

We therefore conclude that the noise spectrum consists of at least two contributions: nuclear spin noise at low frequencies and another mechanism at higher frequencies. At higher frequencies, the noise spectrum decays as $1/f^2$ taking $\alpha=3$, but we see that a $\approx 1/f$ decay (green line) also fits the frequency dependence of the data points well. It is possible that this part of the spectrum is dominated by charge noise, which couples to the spin due to the magnetic field gradient from the micromagnets. Thus charge noise may effectively induce magnetic $1/f$ or $1/f^2$ noise. To give a feeling for numbers, a two-level magnetic field fluctuation of 0.8 $\mu$T, which given the micromagnet gradient corresponds to a $\approx 4$ pm shift back and forth in the dot position, gives a Lorentzian noise spectrum that matches the dashed/solid line in Fig.~3(d) \cite{Bergli2009}.

\section{Randomized benchmarking}

We measured the average gate fidelity using Standard Randomized Benchmarking (SRB), which is known as an efficient way to measure the gate fidelity without suffering from initialization and read-out errors \cite{Knill2008,Magesan2012}. The specific procedure is as follows. After initializing the electron to the spin-down state, we apply randomized sequences of $m$ Clifford gates and a final Clifford gate $C_{m+1}$ that is chosen so that the final target state in the absence of errors is either spin-up or spin-down. Every Clifford gate is implemented by composing $\pi$ and $\pi/2$ rotations around two axes, following \cite{Barends2014}. Applying randomized sequences of imperfect Clifford gates acts as a depolarizing channel \cite{Knill2008,Magesan2012}. The depolarization parameter $p$ reflects the imperfection of the average of 24
 Clifford gates. Under certain assumptions, for $m$ successive Clifford gates the depolarization parameter is $p^m$. 

We measure the spin-up probability both for the case where spin-up is the target state, $P_\uparrow^{\left|\uparrow \right\rangle}$, and for the case where spin-down is the target state, $P_\uparrow^{\left|\downarrow \right\rangle}$, for 119 different randomized sequences for each choice of $m$, and varying $m$ from 2 to 220. The difference of the measured spin-up probability for these two cases, $P_\uparrow^{\left|\uparrow \right\rangle}-P_\uparrow^{\left|\downarrow \right\rangle}$, is plotted with red circles in Fig.~4(a). Theoretically, $P_\uparrow^{\left|\uparrow \right\rangle}-P_\uparrow^{\left|\downarrow \right\rangle}$ is expressed as \cite{Veldhorst2014,Muhonen2015}: 
\begin{eqnarray}
P_\uparrow^{\left|\uparrow \right\rangle}-P_\uparrow^{\left|\downarrow \right\rangle}=ap^m \label{RBeq} 
\end{eqnarray}
where $a$ is a prefactor that does not depend on the gate error. As seen in Eq.~\eqref{RBeq}, differently from quantum process tomography \cite{Kim2014,Chuang2009,Kim2015}, the measurement of the gate fidelity is not affected by the initialization and read-out infidelities, assuming these infidelities are constant throughout the measurement. In order to keep the read-out and initialization fidelities constant for different $m$, we kept the total microwave burst time $t_p+t_p'=150$ $\mu$s. Due to the longer total microwave burst time, the read-out and initialization infidelities are higher than in Fig.~2. This is the reason that initially $P_\uparrow^{\left|\uparrow \right\rangle}-P_\uparrow^{\left|\downarrow \right\rangle}$ is 20\% instead of 45\%. 

Fig.~4(a) shows that the measured decay does not follow a simple exponential $p^m$. This behavior is reproduced by numerical simulations of the randomized benchmarking experiment, using the same set of randomized sequences as used in the experiments and assuming that the magnetic field fluctuations are characterized by $\delta b(t)$, the combination of the high-frequency noise $\delta b'(t)$ and the (quasi-)static noise $\delta b_0$: 
\begin{eqnarray}
\delta b(t)=\delta b_0+\delta b'(t), 
\end{eqnarray}
where $\delta b_0$ again has a Gaussian distribution with FWHM of 0.63 MHz and $\delta b'(t)$ is expressed by Eq.~\eqref{OU_1overf} using $\alpha=3$, $A=2.5 \cdot 10^4$ $\mathrm{rad}^2$s$^{-1}$,  $\tau_c$=2.46, $\mu$s. The simulation results are shown in Fig.~4(b) and show good agreement with experiment.

To evaluate explicitly the relative contribution of $\delta b_0$ and $\delta b'(t)$ to the randomized benchmarking decay, we repeated the numerical simulation including at first only the high-frequency noise $\delta b'(t)$, in which case the decay is extremely slow. Next we include only the (quasi-)static noise $\delta b_0$ and find almost exactly the same decay as with the combination of the two noise contributions. This indicates that the (quasi-)static noise is mainly responsible for the gate error while the contribution from the high-frequency noise is small (SI Appendix). This is consistent with an earlier report \cite{Fogarty2015}, in which it was also shown that ensemble averaging over individual exponential decays can lead to a non-exponential decay. Repeated measurements in the presence of low-frequency noise effectively lead to such ensemble averaging. 

Since the measured SRB decay is not of the form $a p^m$, we should be cautious using the fidelity numbers extracted from this procedure \cite{Epstein2014}. We see that both in experiment and simulation, the decay begins to deviate from a single exponential (straight line in the semi-log plot) for on-resonance microwave bursts with $t_p  \gtrsim 8 \mu$s. These are the data points with open circles in Fig.~4. We fitted to the decay curves for $t_p<8$ $\mu$s to $a p^m$ and obtained $p=0.9620 \pm 0.0051$. From this, the average fidelity of a Clifford gate is $1-(1-p)/2=98.10 \pm 0.26\%$ and the average fidelity for a single $\pi$ or $\pi/2$ rotation around $\hat{x}$ or $\hat{y}$ is calculated to be $1-(1-p)/2/1.875=98.99 \pm 0.14\%$. 

\begin{figure}[!h]
\includegraphics[width=11cm] {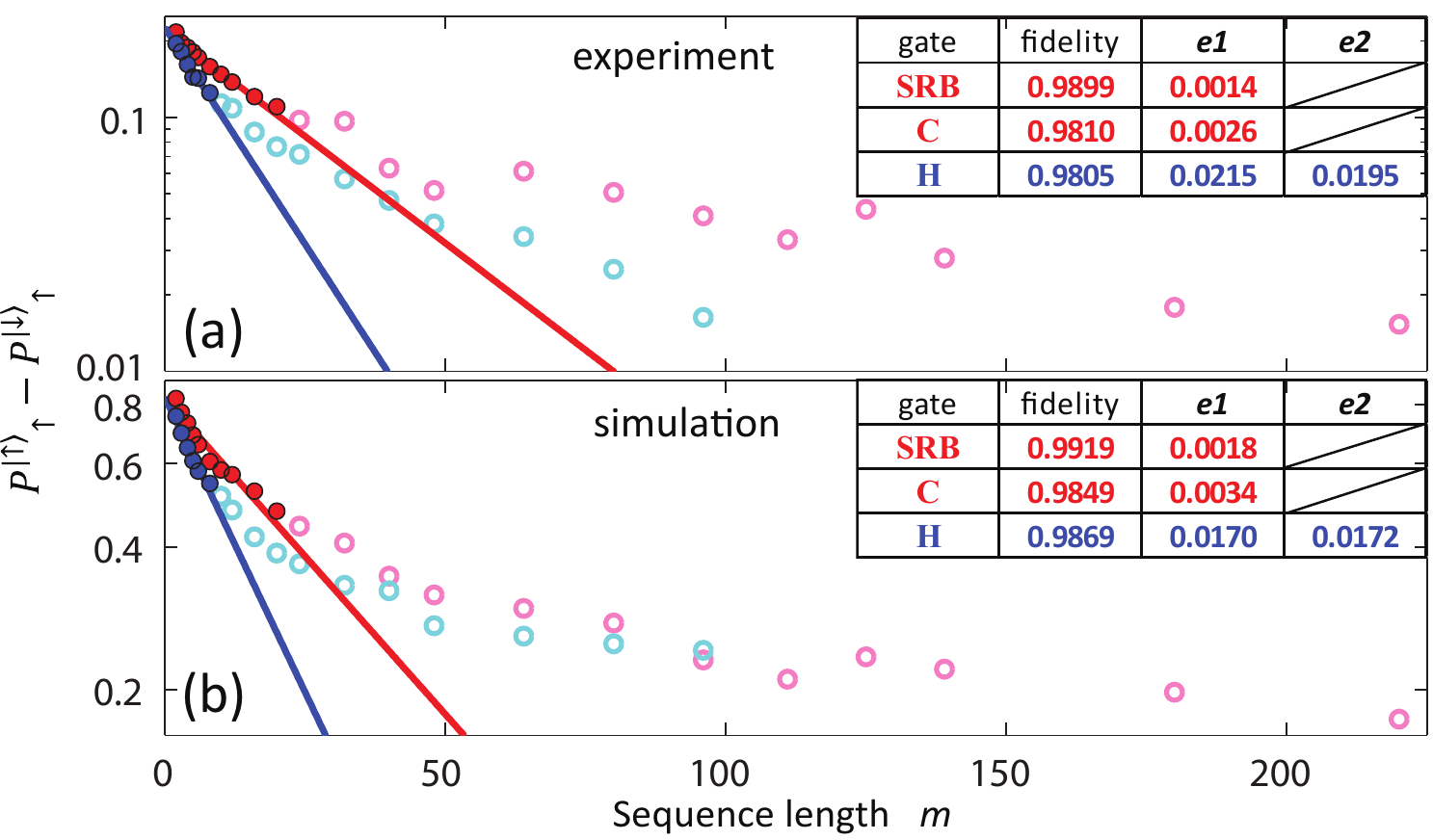}
\caption{\label{fig:fig4} 
Randomized benchmarking.  The difference between the spin-up probability with spin-up as the target state and with spin-down as the target state, $P_\uparrow^{\left|\uparrow \right\rangle}-P_\uparrow^{\left|\downarrow \right\rangle}$, is plotted as a function of the number of Clifford gates, $m$. The Standard Randomized Benchmarking curve (red circles) is measured after applying randomized sequences of $m$ Clifford gates and a final Clifford gate $C_{m+1}$. The Interleaved Randomized Benchmarking curve is measured by interleaving the Hadamard gate with the same random sequence of Clifford gates (blue circles). The experimental results are shown in (a) and the results of numerical simulations (see main text) are shown in (b). The experiments and simulations use the same 119 random sequences from $m=$2 to 220. Each experimental data point is the average of 250 single-shot cycles. For the numerical simulation, we averaged over 1000 repetitions, and for each repetition we sample $\delta b_0$ and include a different time-domain realization of $\delta b'(t)$. In the simulation, the read-out and initialization fidelities are assumed to be perfect. The $\pi$ rotation time is 366 ns for the experiments and 360 ns for the simulation. The delay time between pulses is set to be 5 ns for both the measurements and the simulations. The red and blue curves present fits of the form $A p^m$ to the data with $t_p < 8$ $\mu$s. The gate fidelities extracted from the fits are shown in the insets. The first row (SRB) and the second row (C) show the average fidelity per single gate and per Clifford gate, respectively, obtained from the SRB measurements. The third row (H) shows the fidelity of the Hadamard gate obtained from IRB. $e1$ and $e2$ are as defined in Table 1.}
\end{figure} 

\begin{table}
\includegraphics[width=8cm] {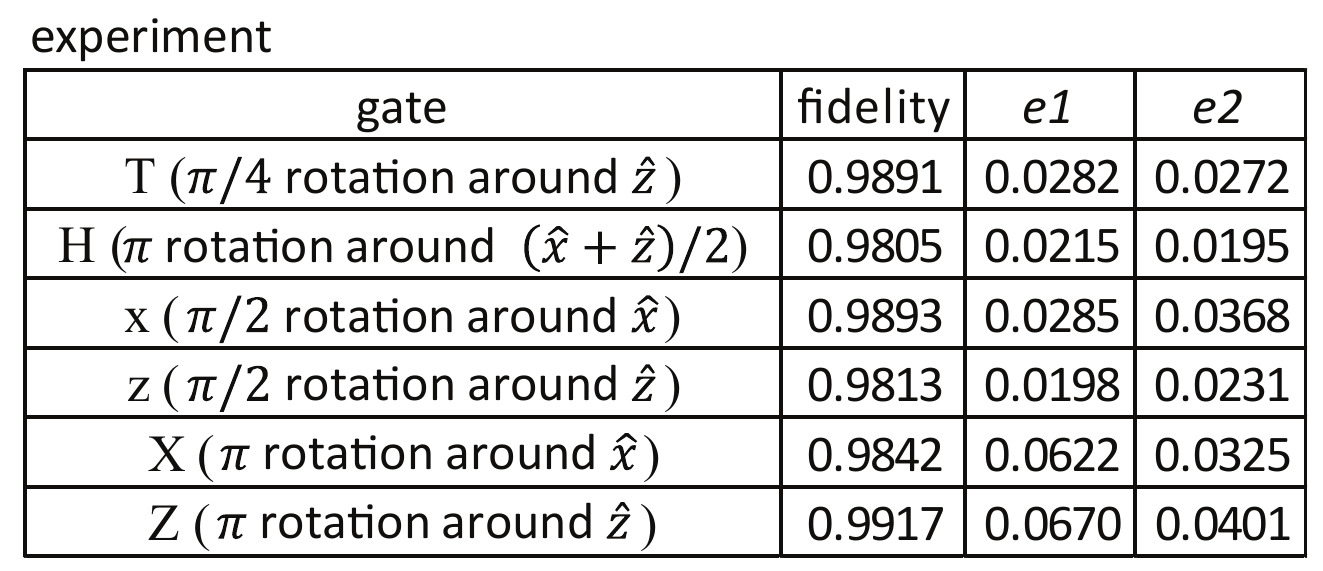}
\caption{\label{tab:tab1} 
The measured gate fidelities for five representative gates, extracted using Interleaved Randomized Benchmarking. $e1$ and $e2$ are errors in the fidelities. $e1$ is calculated from the 95\% confidence interval on the fit coefficient $p$ and $e2$ is an upper bound arising from imperfect random gates,  calculated according to the formulas in \cite{Magesan2012}. As the T gate is not a Clifford gate, we interleaved 2 successive T gates, following \cite{Barends2014}.}
\end{table}

 We also characterized the fidelity of individual gates using interleaved randomized benchmarking (IRB). In this procedure, a specific gate is interleaved between randomized Clifford gates. The depolarizing parameter now becomes bigger than in SRB due to the imperfections of the interleaved gate. From the difference in the depolarizing parameter between SRB and IRB, the fidelity of the interleaved gate is extracted. In Fig.~4(a), the blue circles show the case where the Hadamard gate is the interleaved gate. The Hadamard gate is implemented by a $\pi$ rotation around the $\hat{x}$ axis and a $\pi/2$ rotation around the $-\hat{y}$ axis. By fitting $a_\mathrm{H} {p_\mathrm{H}}^m$ to the decay curve (again for $t_p<8$ $\mu$s), $p_\mathrm{H}=0.9245 \pm 0.0197$ is obtained. The fidelity of the Hadamard gate is calculated to be $1-\left(1-p_\mathrm{H}/p\right)/2=98.05 \pm 2.15\%$. In the same way, we measured the fidelities for several other common gates (Table~1). While also for IRB, the decay is not exponential, the gate fidelities extracted from IRB for the first 8 $\mu$s appear roughly consistent with the fidelities extracted from SRB.
 
\section{Discussion and Conclusion}
We have shown that the average single gate fidelity for a single electron spin confined in a $^\mathrm{nat}$Si/SiGe quantum dot approaches the fault-tolerance threshold for surface codes \cite{Fowler2012}. The low frequency noise that limits gate fidelity is well explained by the nuclear spin randomness given the natural abundance of $^\mathrm{29}$Si. Therefore we can increase gate fidelities by reducing the abundance of $^\mathrm{29}$Si using isotopically enriched $^\mathrm{28}$Si \cite{Veldhorst2014,Muhonen2014} or by using composite pulses \cite{Vandersypen2005}.
The longest coherence time measured using dynamical decoupling is $\approx$ 400 $\mu$s. We revealed that the noise level is flat in the range of 5 kHz - 30 kHz and decreases with frequency in the range of 30 kHz - 1 MHz. In this frequency range (5 kHz - 1 MHz), the measured noise level is higher than expected from the dynamics of the $^\mathrm{29}$Si nuclear spins. Instead, charge noise in combination with a local magnetic field gradient may be responsible. If charge noise is dominant, dynamical decoupling decay times can be further extended by positioning the electron spin so that the gradient of the longitudinal component of the magnetic field gradient vanishes, while keeping the transverse component non-zero as needed for driving spin rotations. At that point, we can reap the full benefits from moving to $^\mathrm{28}$Si enriched material for maximal coherence times as well.

\section{ }
\begin{acknowledgments}
We acknowledge R. Hanson, G. de Lange, M. Veldhorst, S. Bartlett and L. Schreiber for useful discussions, and R. Schouten and R. Vermeulen for technical support. This work was supported in part by ARO (W911NF-12-0607), FOM and a ERC Synergy grant; development and maintenance of the growth facilities used for fabricating samples is supported by DOE (DE-FG02-03ER46028). E.K. was supported by a fellowship from the Nakajima Foundation. This research utilized NSF-supported shared facilities at the University of Wisconsin-Madison. Work at the Ames Laboratory (analysis of nuclear spin noise and decoherence) was supported by the Department of Energy-Basic Energy Sciences under Contract No. DE-AC02-07CH11358.
\end{acknowledgments}
\newpage
\section{Supplementary materials}
\appendix

\renewcommand{\thefigure}{S\arabic{figure}}
\renewcommand{\theequation}{S\arabic{equation}}

\section{Modeling of decoherence and noise produced by nuclear spins}

In order to analyze decoherence and noise produced by the nuclear spins, we performed numerical modeling of the Hahn echo decay and of the spectrum of the nuclear spin noise using the coupled cluster expansion (CCE) \cite{Yang2009} within second order (i.e., considering the nuclear spin pairs). To model the quantum dot we generate a crystallite of silicon by representing the underlying diamond lattice as two fcc sublattices shifted with respect to each other by the vector [1/4,1/4,1/4]. The size of the crystallite is $(2L_x+1)\times (2L_y+1)\times (2L_z+1)$ along the $x$-, $y$-, and $z$-axis, respectively. For simulation results shown below we used the external field directed along the $x$-axis as in the experiments and $L_x=L_y=18$ and $L_z=8$ which is large enough, so that the results do not depend much on the crystallite size; we also did not see any significant changes with changing the orientation of the dot with respect to the external field directed along the $z$-axis. The sites inside the crystallite are randomly populated with spins 1/2 (which represent $^{29}$Si nuclei) with abundance of 4.68\%, so the total number of nuclear spins in the crystallite was about 8700. For convenience, below we express the coordinates $(x,y,z)$ in the units of $d$, where $d=0.543$~nm is the lattice constant of the cubic lattice.

The envelope of the electron density within the dot is modeled as a 3-dimensional Gaussian distribution, centered at the central unit cell of the crystallite:
\begin{equation}
\rho_{env}(x,y,z) = \rho_0 \exp \left( \frac{-x^2}{2 x_0^2}\right) \exp \left( \frac{-y^2}{2 y_0^2} \right) \exp \left( \frac{-z^2}{2 z_0^2} \right),
\end{equation}
with various deviations $x_0$, $y_0$, and $z_0$. Below, we present the results for two representative cases: (i) $x_0=0.4\cdot L_x$, $y_0=0.4\cdot L_y$, and $z_0=0.4\cdot L_z$, denoted as the larger dot, and (ii) $x_0=0.2\cdot L_x$, $y_0=0.2\cdot L_y$, and $z_0=0.2\cdot L_z$, denoted as the smaller dot. The hyperfine coupling constants are proportional to the electron density, so that the Hamiltonian of the hyperfine coupling between nuclear spins and the electron spin is
\begin{equation}
H =\sum_{i=1}^{N_n} A_i I_{ix} S_x 
\end{equation}
with $A_i=A_0 \rho_{env}(x_i,y_i,z_i)$, where $x_i$, $y_i$, and $z_i$ are the coordinates of the $i$-th nucleus, its spin operator is $I_{ix}$, while $S_x$ is the electron spin operator (note that the external field is directed along the x-axis), and $N_n \sim 8700$ is the total number of $^{29}$Si nuclear spins in the crystallite. The proportionality coefficient $A_0$ is normalized to produce the standard deviation in electron Larmor frequency $\sigma_\omega=\sqrt{\sum_i {A_i}^2 }/2$ equal to 1.7 Mrad/s, i.e. approximately $2\pi \cdot 271$~kHz, which corresponds to the measured line width (or, equivalently, to $T_2^*=800$ $\mu$s or to a random static field of 9.6~$\mu$T). Calculated value of $A_0$ for a larger dot is almost consistent with a theoretically calculated $A_0$ for natural silicon \cite{Assali2011}.

The nuclear spins are subjected to an external field $B_\mathrm{ext}=800$~mT and a gradient field from the micromagnet. The gradient field is much smaller than $B_\mathrm{ext}$, so that the quantization axes of different nuclear spins remain very close to the $x$-axis, and only the magnitude of the local field varies from one nucleus to another. Correspondingly, the Zeeman part of the $i$-th nuclear spin Hamiltonian is
\begin{equation}
H_{Z,i}= \gamma_n\hbar\ I_{ix} \ B(x_i)
\end{equation}
where $\gamma_n$ is the gyromagnetic ratio of the $^{29}$Si nucleus. We only consider the gradient of the $x$ component of the micromagnet's field along the $x$ axis, i.e., parallel to the external magnetic field, so that $B(x_i)=B_\mathrm{ext}+x_i G_\parallel $, where $G_\parallel$ is the gradient of the $x$ component of the stray magnetic field, where $\gamma_n \hbar G_\parallel=5.78$~krad/s (approximately $2\pi \cdot 0.92$~kHz) per unit cell. Note that the variations in the direction of the local quantization axes can be easily taken into account, but their effect is negligible, leading only to small renormalization of the hyperfine coupling constants $A_i$ and of the dipole-dipole couplings between the nuclear spins, and the direction of the quantization $z$-axis being different for different nuclear sites. 

\begin{figure}[h!]
\centering
\includegraphics[width=13cm]{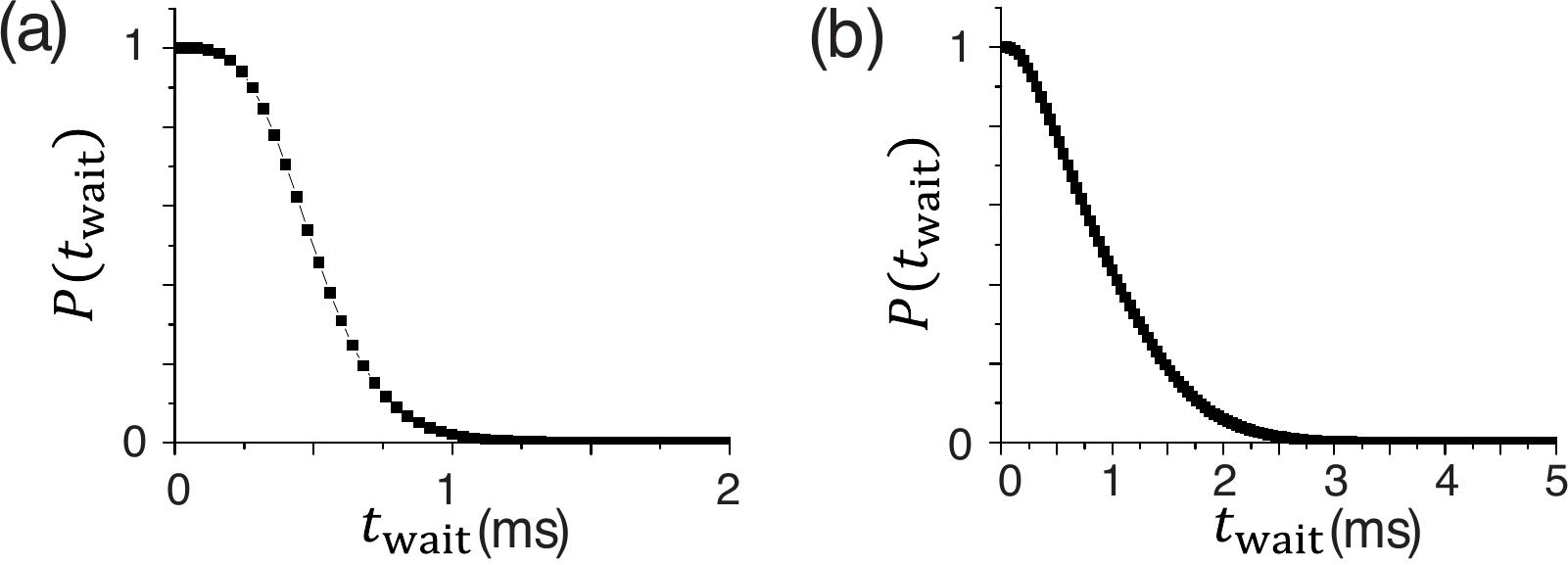}
\caption{Hahn Echo amplitudes $P(t_\mathrm{wait})$ for the larger dot (a) and the smaller dot (b) as a function of the total waiting time $t_\mathrm{wait}$.}
\label{nucspinsEcho}
\end{figure}

The dipolar interaction between the nuclear spins is given by the sum of interaction between all possible pairs
\begin{equation}
H_{nn} = \sum _{i<j} b_{ij} (1-3 \cos^2 \theta_{ij} ) \left( I_{ix} I_{jx} -\frac{1}{2} I_{iz} I_{jz} - \frac{1}{2} I_{iy} I_{jy} \right)
\end{equation}
with $b_{ij}=\frac{ \gamma_n^2 \hbar}{R^3_{ij}} $, where $R_{ij}=r_{ij}d$ is the distance between the two nuclei, and $\theta_{ij}$ is the angle between the line connecting the nuclei and the $x$-axis. We have also included the electron spin-mediated coupling between the nuclear spins \cite{Cywinski2009} in the numerical modeling, of the form 
\begin{equation}
H^{(2)}_{nn}=\sum _{i<j} \beta_{ij} S_z \left( \frac{1}{2} \left( I_{iz} I_{jz} + I_{iy} I_{jy} \right) \right)
\end{equation}
with $\beta_{ij}=\frac{A_i A_j}{2\omega_0}$ with $\omega_0$ the Larmor frequency of the electron spin. As expected, this coupling did not produce any noticeable effect, since the magnitude of the electron spin-mediated coupling is much smaller than the direct dipolar interaction between nuclei. Indeed we can roughly estimate the typical strength of the dipolar interaction between nuclear spins as $\gamma_n^2\hbar {\bar R}^{-3} \approx 66$~rad/s where $\bar{R}^{-3}$ is the density of nuclear spins, while the typical electron spin-mediated coupling is of the order of $A_k^2/(2\omega_0) \sim 2 \sigma_w /(N_s \omega_0)$, which is about 0.1-0.01 rad/s for $B_\mathrm{ext} \sim 500$~mT; here $N_s$ is the number of the nuclear spins effectively coupled to the electron, which is about 4500 for the larger dot and 560 for the smaller dot so that the magnitude of the electron spin-mediated coupling is several orders smaller than the dipolar coupling for such a large magnetic field.  As expected form this estimation, the electron spin-mediated coupling did not produce a noticeable effect in the calculation result. The Hahn echo decays calculated within the pair-cluster CCE are shown in Fig.~\ref{nucspinsEcho} for the larger and the smaller quantum dots (defined above). The Hahn echo decay times are of order of 0.5--1~ms, which is much longer than the experimentally measured $T_2$ time. Similar results have been obtained for other system parameters we explored. 

In order to gain deeper insight into the nature of the noise created by the nuclear spins, we used the pair-cluster CCE to calculate the correlator of the nuclear spin noise. Let us first note that the concept of the nuclear spin noise, which acts on the electron spin and decoheres it, is a semi-classical concept. For instance, it presumes that the properties of the nuclear spin noise (such as the correlation function) are well defined and independent of the electron spin. However, this description is not always adequate: much decoherence comes from the quantum back-action of the electron spin on the nuclear bath, so that the motion of the nuclear spins depends on the electron spin state. Nevertheless, the model of the random nuclear noise has its advantages, and often gives a reasonable semi-quantitative description of decoherence for a range of experimentally interesting situations \cite{Reinhard2012}.

\begin{figure}[h!]
\centering
\includegraphics[width=15cm]{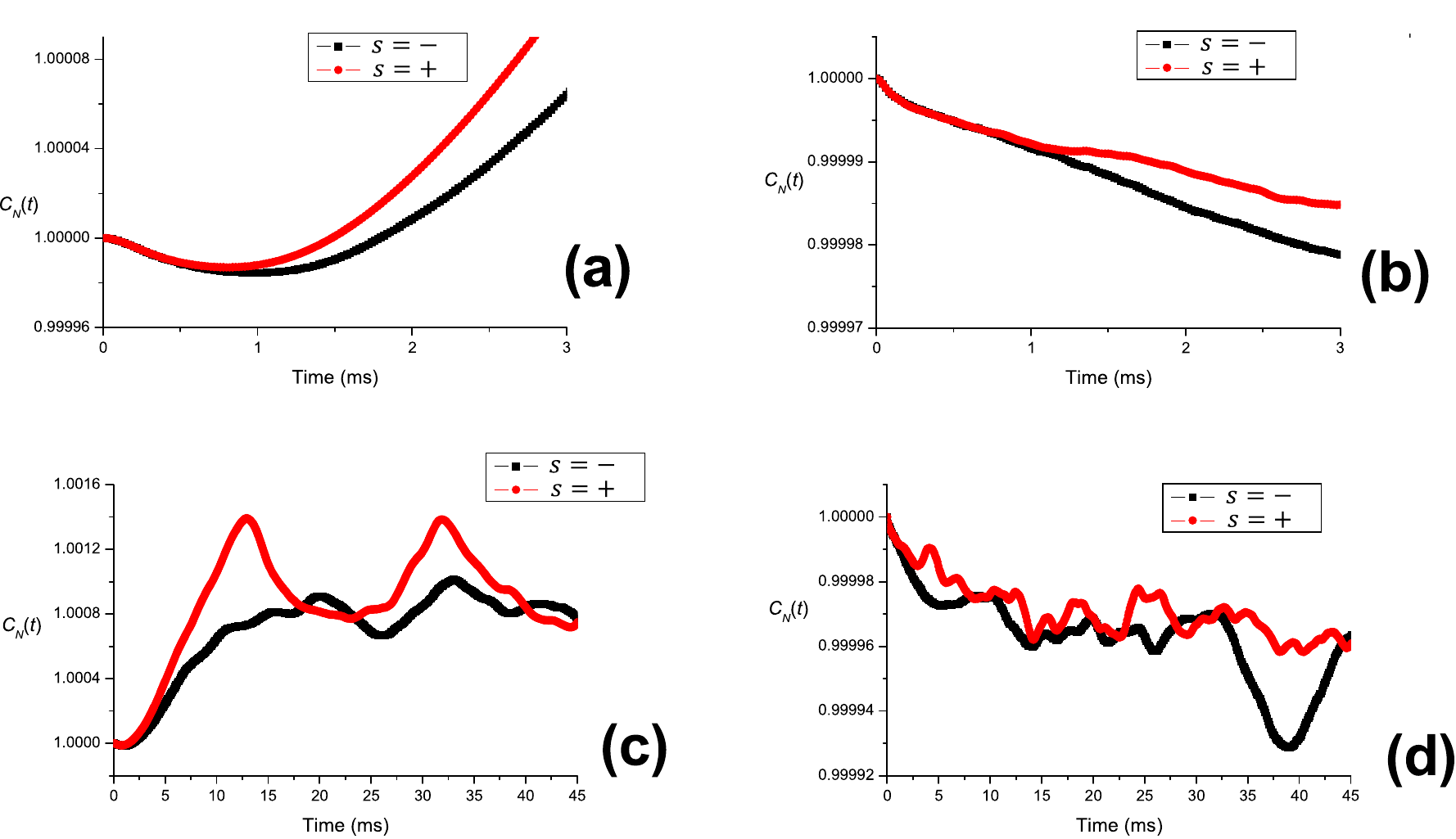}
\caption{Normalized correlation functions $C_N(t)$ for different orientations of the electron spin, for the two quantum dots: (a) and (c) --- for the larger quantum dots; (b) and (d) --- for the smaller quantum dot. Panels (a) and (c) show the same results, but for different time ranges, the same is true for the panels (b) and (d).}
\label{nucspinsCorr}
\end{figure}

Thus, we investigate the correlation function of the spin noise. Specifically, for a given state of the electron spin $s=+,-$ (along the x-axis or in the direction opposite to the x-axis) we calculate the correlator
\begin{equation}
C(t)=\langle \eta_s(0) \eta_s(t)\rangle = \sum_{i,k} A_i A_k \langle I_{ix} I_{kx}(t)\rangle_s
\end{equation}
where $\langle\dots\rangle_s$ denotes the quantum-mechanical average with the completely disordered initial state of the nuclear spin bath and the electron spin in the state $s$ and $\eta(t)=\sum_i^{N_n} A_i I_{ix}(t) =\gamma_e \delta b(t)$ with $\gamma_e $ gyro-magnetic ratio of the electron and $\delta b(t)$ effective magnetic field fluctuation along the z axis due to the nuclear spin bath. We performed simulations with the pair-cluster CCE method, using the Heisenberg representation for the nuclear spin operators $I_{mx}(t)$. Since the correlation function $C(t)$ has the value $C(0)={\sigma_\omega}^2$ at $t=0$ (where $\sigma_\omega=2\pi \cdot 271$~kHz), it is convenient to plot the normalized correlator $C_N(t)=C(t)/{\sigma_\omega}^2$; these simulation results are plotted in Fig.~\ref{nucspinsCorr}.

Our results show that the noise correlators corresponding to different electron spin states can be very different, implying back-action of the electron spin on the nuclear spin bath. Indeed, the hyperfine coupling constants, even for quite large quantum dots, are large in comparison with the dipole-dipole couplings between the nuclear spins. However, the echo decay is determined by the noise dynamics {\it at the timescale of the echo decay\/}, which is of the order of a few milliseconds. At this timescale the noise correlators corresponding to different electron spin states do not differ too much, and give meaningful semi-quantitative information about the properties of the nuclear spin bath.

Moreover, let us note that although the overall behavior of the correlation functions on the timescales of order of 10--50~ms is quite irregular, the experiments probe only much smaller timescales, of order of few milliseconds. 
Since the internal dynamics of the nuclear spin bath is very slow, the overall change in the noise amplitude over this timescale is very small. For better understanding, let us compare the free decay of the electron spin coherence and the Hahn echo decay assuming that the nuclear spin noise is described as an Ornstein-Uhlenbeck (Gaussian, Markovian, stationary) random process \cite{Wang2012} with amplitude $\sigma_\omega$ and correlation time $\tau_c$. The free decay time $T_2^*$ is inversely proportional to the noise amplitude $\sigma_\omega$, while the Hahn echo decay time $T_2$ is proportional to $(b^2/\tau_c)^{-1/3}$, so that
\begin{equation}
\frac{\tau_c}{T_2^*} \sim \left(\frac{T_2}{T_2^*}\right)^3,
\end{equation}
i.e.\ for the quantum dot under consideration, the correlation time of the nuclear noise is of order of $2\cdot 10^8$ times longer than $T_2^*$. I.e., on the timescale of the Hahn echo decay, the change in the correlation function $\langle \eta_s(0) \eta_s(t)\rangle$ is miniscule, see also Fig. S2.

\begin{figure}[h!]
\centering
\includegraphics[width=13cm]{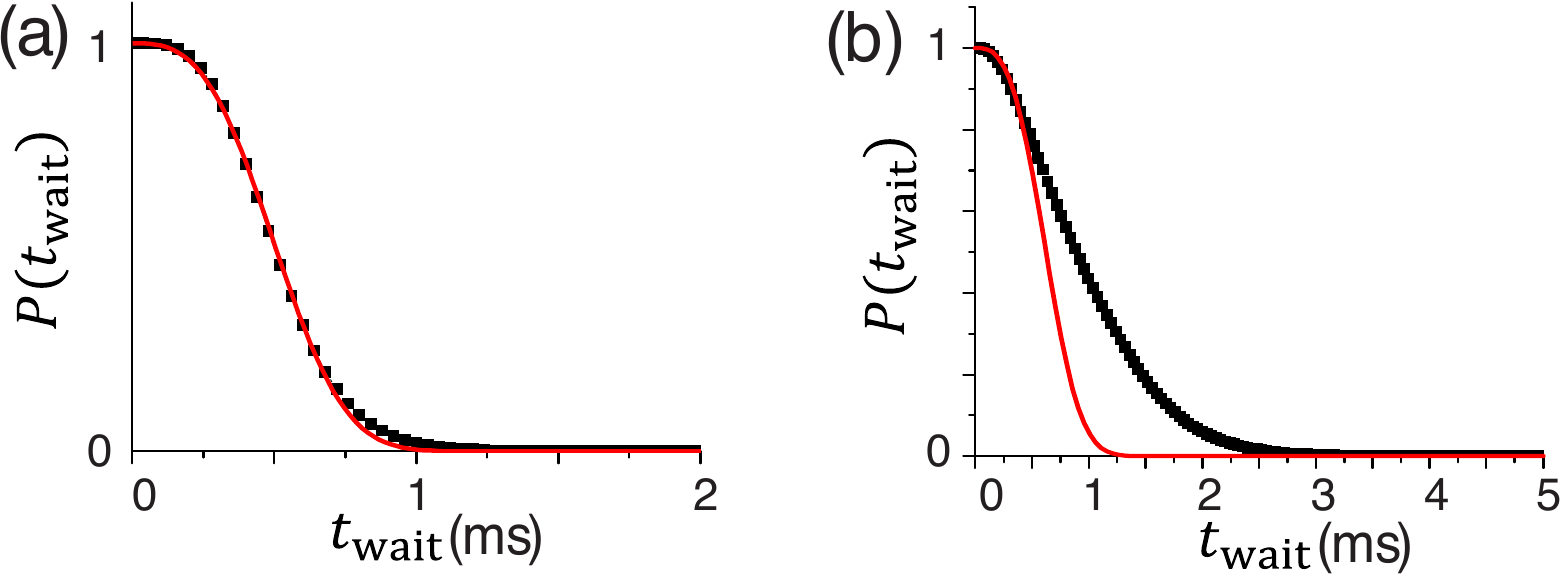}
\caption{Calculated Hahn echo amplitudes (black squares) for the larger dot (a) and for the smaller dot (b), obtained from the CCE simulations, are compared with the theoretical estimates (Eq.~\ref{ecestimate}): shown as red lines.}
\label{nucspinsComp}
\end{figure}

In the experimentally relevant region of times, $<$1~ms, the calculated correlation functions can be approximated by decaying exponentials $C_N(t)=\exp{(-t/\tau_c)}$. From such a fitting, we extract the values of $\tau_c$ and substitute it into the simplified theoretical expression for the Hahn echo decay
\begin{equation}
P(t_\mathrm{time})=\exp{\left(-  \frac{t_\mathrm{wait}^3 \sigma_\omega ^2}{ 12 \tau_c}   \right)}.
\label{ecestimate}
\end{equation}
Fig.~\ref{nucspinsComp} illustrates the comparison between such estimates and the original CCE-simulated echo curves (shown above in Fig.~\ref{nucspinsEcho}). The values of $\tau_c$ were obtained from fitting the curves $C_N(t)$ to the decaying exponent in the region of $0<t<0.5$~ms. This is the range of times which controls the echo decay, and where the correlators for the electron spin states "down" ($s=-$) and "up" ($s=+$) are close to each other. The resulting values are $\tau_c=2.24\cdot 10^{-5}$~ms for the larger dot, and $\tau_c=1.18\cdot 10^{-5}$~ms for the smaller dot. The red curves in Fig. S3 describe well the echo decay during the first millisecond, where the approximation of the nuclear noise is valid, and where the back-action from the electron spin is not too strong.

Thus, we conclude that the model of the nuclear spin noise is semi-quantiatively applicable to the considered quantum dots. However, the experimentally measured Hahn echo decay time is an order of magnitude shorter than the CCE simulations predict for the nuclear spin-induced echo decay. Similarly, we can extract the noise spectrum using the exponential fittings of $C(t)$ described above: the result is shown in Fig.~3(d) of the main text. Again, we see that the noise produced by the nuclear spins is too slow in comparison with the experimentally measured one, and is unlikely to be an important source of decoherence in the Hahn echo and the dynamical decoupling experiments.

\section{Heating of the electron reservoir as a function of microwave burst time}
\begin{figure}[h!]
\centering
\includegraphics[width=15cm] {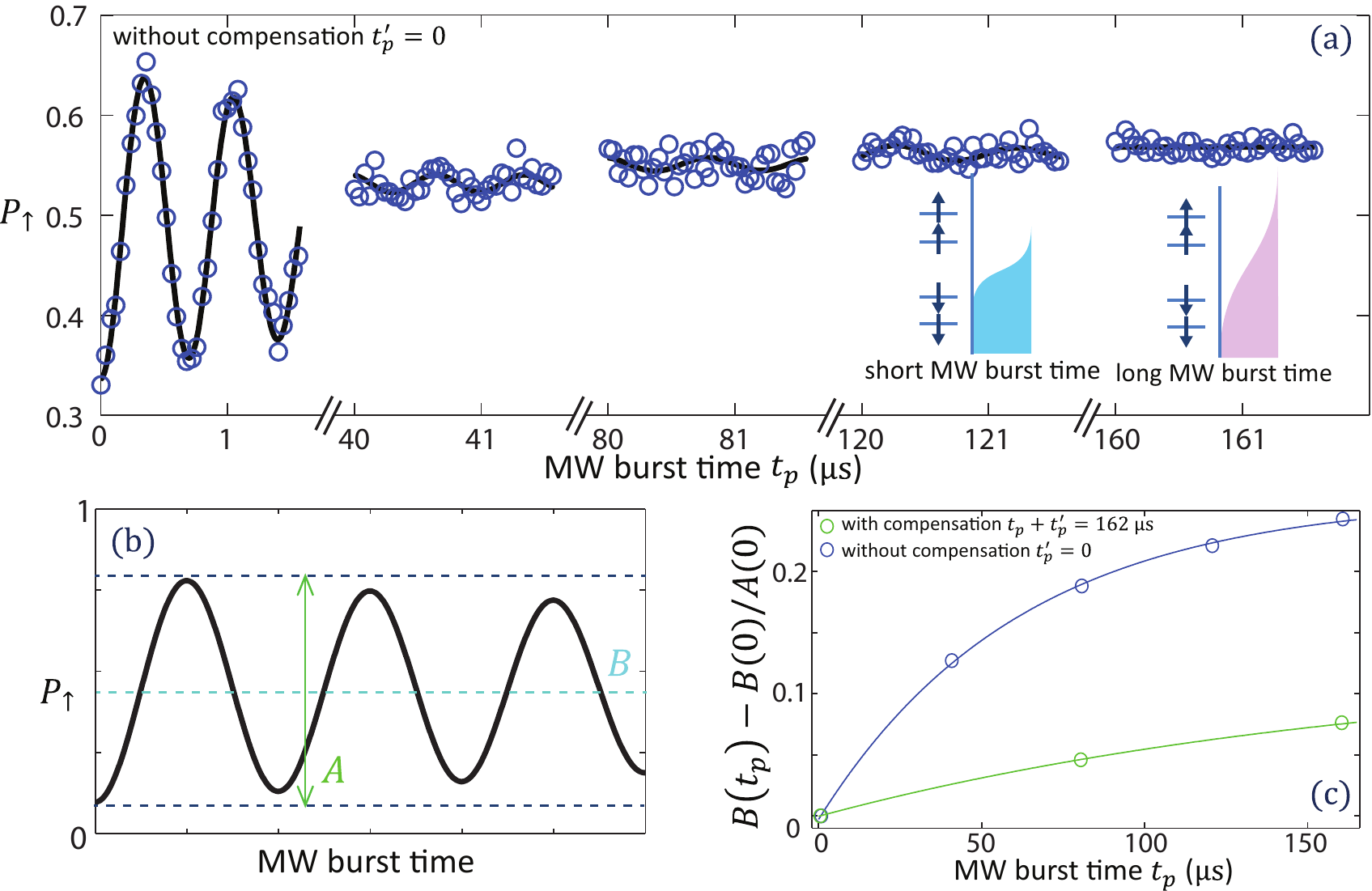}
\caption{\label{fig:OffsetIncreasingRabi} (a) Measured spin-up probability, $P_\uparrow$, showing a Rabi oscillation for the ground valley-orbit state (blue circles) without off-resonance microwaves applied. The black lines show fits with damped cosine curves assuming that the offset, $B$, can be regarded constant for short time (1.5 $\mu$s).  The insets show schematics of the four energy levels (two spin states and two valley states) in the dot and the electron reservoir for a short microwave burst time and for a long microwave burst time. (b) The offset $B$ and the amplitude $A$ in the measured spin-up probability of a Rabi oscillation. The deviation of the amplitude $A$ from 1 and that of the offset $B$ from 0.5 are due to the initialization and read-out infidelities. (c) The increase of the offset $B(t_p)-B(0)$ normalized by the initial amplitude $A(0)$ as a function of the microwave burst time $t_p$, without off-resonance microwaves (blue circles), and with off-resonance microwaves (green circles) so that the duration of the microwave bursts is fixed to 162 $\mu$s.}
\end{figure} 
Fig.~\ref{fig:OffsetIncreasingRabi}(a) shows a Rabi oscillation of the ground valley-orbit state varying the microwave burst time up to $\approx$ 160 $\mu$s. We take the offset of a Rabi oscillation in the measured spin-up probability as $B$ and its amplitude as $A$ as shown in Fig.~\ref{fig:OffsetIncreasingRabi}(b). The offset $B$ increases with longer microwave burst time, while in Fig. S4(a) there is no change in the amplitude $A$ other than the decrease expected by quasi-static noise, is observed. These results indicate that the microwave bursts heat the electron reservoir.

To quantify this effect, we characterize the spin read-out and initialization fidelities of the excited valley-orbit state (ground valley-orbit state) by three parameters, $\alpha_1$, $\beta_1$ and $\gamma_1$ ($\alpha_2$, $\beta_2$ and $\gamma_2$) following the supplementary materials of \cite{Kawakami2014}. By taking the population of two valley-orbit states as $\epsilon_1:\epsilon_2$, the amplitude, $A$, and the offset, $B$, are expressed with 
\begin{eqnarray}
A=\epsilon_2 (1-2\gamma_2)(1- \beta_2- \alpha_2)
\end{eqnarray}
and
\begin{eqnarray}
B=B_1+B_2,
\end{eqnarray}
with
\begin{eqnarray}
B_1=\epsilon_1 \left[ (1-\beta_1) \gamma_1 +\alpha_1 (1-\gamma_1) \right]
\end{eqnarray}
and
\begin{eqnarray}
B_2=\epsilon_2 \frac{1-\beta_2+\alpha_2}{2}.
\end{eqnarray}
$A$ is determined only by the parameters for the ground valley-orbit state. Since no change in $A$ is observed, we conclude that the read-out and initialization fidelities for the ground valley-orbit state are not significantly affected by increased burst times. For the same reason, $B_2$ should not greatly change. Thus the increase in $B_1$ is the dominant source for the increase in $B$. As shown in the insets of Fig.~\ref{fig:OffsetIncreasingRabi}, we assume that the electron temperature of the reservoir is increased with longer microwave burst time. The spin-down excited valley-orbit state is the closest dot energy level to the Fermi level of the reservoir and thus is affected the most by the increase of the electron temperature. 

Fig.~\ref{fig:OffsetIncreasingRabi}(c) shows the change of the offset, $B(t_p)-B(0)$, normalized by the initial amplitude, $A(0)$, as a function of the microwave burst time. The blue circles are extracted from Fig.~\ref{fig:OffsetIncreasingRabi}(a) and green circles are extracted from a similar measurement as in (a) but with off-resonance microwave excitation applied in order to keep the total duration of the microwave bursts fixed to 162 $\mu$s. The increase of the offset is heavily suppressed by applying the off-resonance microwave but not completely suppressed. This remaining increase may be due to a change in the population between the two valley-orbit states with long microwave burst time. With the total microwave time $<8$ $\mu$s (condition assumed for the evaluation of the gate fidelities with randomized benchmarking), this remaining increase is small enough to be ignored. Therefore the measured gate fidelities are not affected by this effect. Also for the dynamical decoupling measurement, the longest total microwave duration used is $\approx 8$ $\mu$s and we have not observed any significant difference in the dynamical decoupling coherence time with versus without off-resonance microwave excitation as discussed below.

\section{Extract the noise spectral density from the echo decays}
Here we show how the noise spectral density is extracted from the experimental echo decay data. When a dynamical decoupling pulse sequence whose time separation between $\pi$ pulses is fixed and symmetric i.e. the timing of $k$th $\pi$ pulse is $t_k=\frac{t_\mathrm{wait}}{N_\pi}$), the filter function $F(\omega,t_\mathrm{wait},N_\pi)$ peaks at $\omega_0=\frac{2\pi }{4 \tau}$ with $\tau=\frac{t_\mathrm{wait}}{2 N_\pi}\left(k-\frac{1}{2} \right)$.

Following Bylander et al. \cite{Bylander2011}, if the filter function $F(\omega)$ is sufficiently narrow around $\omega_0$, we can treat the noise as constant within the bandwidth of the filter function and then Eq.~(2) is reduced to
\begin{eqnarray}
W(t_\mathrm{wait}) \approx S(\omega_0) \int_{-\infty}^{\infty} \frac{1}{2\pi} \frac{F(\omega,t_\mathrm{wait},N_\pi)}{\omega^2} d\omega. \label{WApprox} 
\end{eqnarray}
According to the numerical simulation we confirmed that Eq.~\eqref{WApprox} is valid for $N\geq 4$.
From Eq.~(1) we know that 
\begin{eqnarray}
W(t_\mathrm{wait})=-\log P\left(t_\mathrm{wait}\right) \label{PtoW} 
\end{eqnarray}
and so $S(\omega_0)$ is determined by the logarithm of the normalized echo amplitude at time $t_\mathrm{wait}=2 \tau N_\pi$ divided by the integral of the filter function:
\begin{eqnarray}
S\left(\omega_0=\frac{2\pi}{4 \tau} \right) \approx \frac{-\log P(t_\mathrm{wait})}{\int_{-\infty}^{\infty} \frac{1}{2\pi} \frac{F(\omega)}{\omega^2} d\omega} = \frac{-2 \log P(t_\mathrm{wait})}{t_\mathrm{wait}}. \label{SApprox} 
\end{eqnarray}

\section{Further analysis of the echo decays}
\begin{figure}[h!]
\centering
\includegraphics[width=\textwidth]{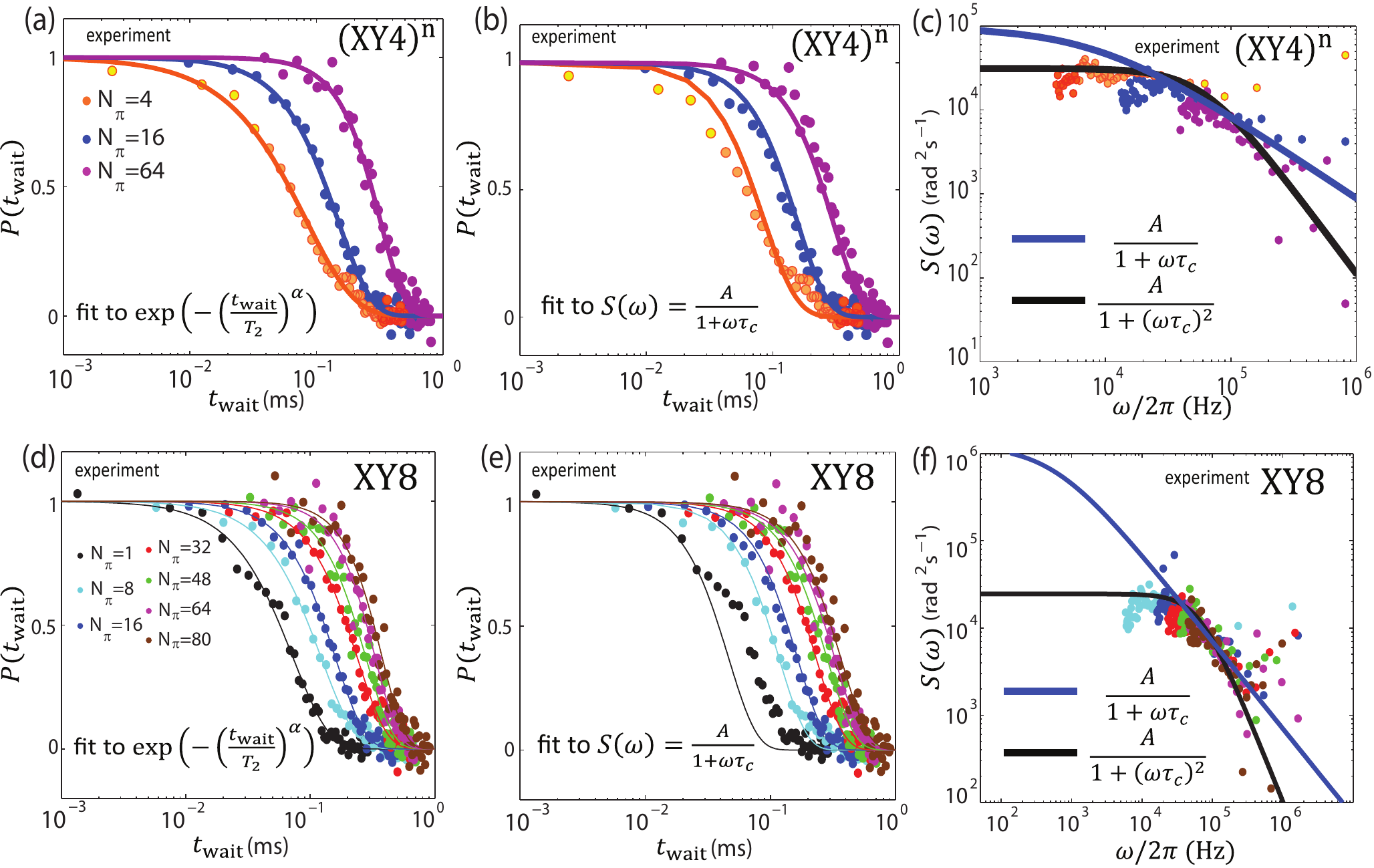}
\caption{\label{fig:vCDDfitOU1overffit1overf} Results of different fits of echo decay curves for (XY4)$^n$ and XY8 pulse sequences. The circles in (a,b) and in (d,e) show the echo decay curves using (XY4)$^n$ and XY8 pulse sequences, the same data as in Fig.~3(a) and (b), respectively. (a,d) The solid lines are the fits using Eq.~(4) with $\alpha$ and $T_2$ as fitting parameters. The experimental data shown with circles in (a-d) are normalized according to these fits. (b,e) The solid lines are the numerical fits using Eq.~(1) with Eq.~(2) and Eq.~(5) for $\alpha=2$. See the text for the fitting results. (c,f) The circles show the noise spectrum extracted from the experimental data of (XY4)$^n$ and XY8, respectively. The blue solid lines present Eq.~(5) for $\alpha=2$ with $A$ and $\tau_c$ obtained from the fits in (b.e). The black solid lines present Eq.~(5) for $\alpha=3$ with $A$ and $\tau_c$ obtained from the fits show in Fig.~3(a,b). (a-c) The decay curve for $N_\pi=4$ and the noise spectrum extracted from this decay curve are showed with 3 different colored circles: yellow, orange and strong orange for short, middle and long waiting time, respectively, in order to demonstrate which parts of decay data correspond to the extracted spectrum noise.
}
\end{figure} 
 The circles in Fig.~\ref{fig:vCDDfitOU1overffit1overf}(a,b) and (d,e) show the echo decay curves using (XY4)$^n$ and XY8 pulse sequences, respectively (the same data as in Fig.~3(a)). In the main text, we fitted the echo data to Eq.~(1) with Eq.~(2) and Eq.~(5) keeping $\alpha$ fixed to 3. Here, we explore alternative ways of fittings.
 The solid lines in Fig.~\ref{fig:vCDDfitOU1overffit1overf}(a,d) are fits using Eq.~(4) with $\alpha$ and $T_2$ as fitting parameters. In Fig.~\ref{fig:vCDDfitOU1overffit1overf}(a), $\alpha=1.14$, 1.64 and 2.22 and $T_2=83.6$, 149 and 320 $\mu$s are obtained for $N_\pi=4$, 16 and 64, respectively. In Fig.~\ref{fig:vCDDfitOU1overffit1overf}(d), $\alpha= $1.50, 1.59, 1.82, 1.85, 2.03, 2.31 and 2.38 and $T_2=72.7$, 122, 159, 239, 280, 333 and 378 $\mu$s are obtained for $N_\pi=1$, 8, 16, 32, 48, 64 and 80 respectively. $\alpha$ becomes larger with higher numbers of $\pi$ pulses because the echo decay curve becomes determined more by the decaying parts of the noise spectrum rather than the flat part.

 The solid lines in Fig.~\ref{fig:vCDDfitOU1overffit1overf}(b,e) are fits using Eq.~(1) with a numerical integration of Eq.~(2) and Eq.~(5) for $\alpha=2$. The fits yield $A=(9.8 \pm 4.7) \cdot 10^4$ $\mathrm{rad}^2$s$^{-1}$ in (b) and $\tau_c=17.4 \pm 8.29$ $\mu$s and $A=(124.5\pm 827) \cdot 10^4$ $\mathrm{rad}^2$s$^{-1}$ and $\tau_c=280 \pm 18.7$ $\mu$s in (e). 

 The circles in Fig.~\ref{fig:vCDDfitOU1overffit1overf}(c) and (f) show the noise spectrum extracted from the experimental data of (XY4)$^n$ and XY8, respectively. The blue solid lines in Fig.~\ref{fig:vCDDfitOU1overffit1overf}(c) and (f) present Eq.~(5) for $\alpha=2$ with $A$ and $\tau_c$ obtained from fitting the decay curves in Fig.~\ref{fig:vCDDfitOU1overffit1overf}(b) and (e), respectively. The black solid line in Fig.~\ref{fig:vCDDfitOU1overffit1overf}(c) (Fig.~\ref{fig:vCDDfitOU1overffit1overf}(f)) shows Eq.~(5) for $\alpha=3$ with $A=2.5\pm 0.2 \cdot 10^4$ $\mathrm{rad}^2$s$^{-1}$ and $\tau_c=2.46 \pm 0.17$ $\mu$s ($A=3.1\pm 0.2 \cdot 10^4$ $\mathrm{rad}^2$s$^{-1}$ and $\tau_c=2.64 \pm 0.19$ $\mu$s), which are obtained from the fitting of the decay curves using the analytical expression as shown in Fig.~3(a) (Fig.~3(b)). Furthermore we found that the result of the fitting using the numerical integration of Eq.~(5) for $\alpha=3$ coincides with that of the the fit using the analytical expression. We conclude that Eq.~(5) with $\alpha=3$ (Lorentian spectrum) captures the echo decay data for a wider frequency range than with $\alpha=2$.
%

\section{Generation of noise for the numerical simulations}
In this section, we explain how the noise fluctuation, $\beta(t)=\gamma_e \delta b'(t)$, which exhibits a spectral density $S(\omega)$, is generated. We assume that $\beta(t)$ is constant for a short time $\Delta t$.
We first generate $u^{\alpha}(t_i)$, discrete points of a noisy time trace corresponding to a white noise spectrum, with the mean $E=0$ and the standard deviation $\sigma_0=1$, where $\alpha$ indexes a different set of noise sequence (different set of measurement) and $t_i=i \Delta t$ ($i=1,2,...,N$) using the Mathematica function RandomVariante[NormalDistribution[$E$,$\sigma_0$]].

The discrete Fourier transform of $u^{\alpha}(t_i)$ is defined as 
\begin{eqnarray}
\tilde{u}^{\alpha}(\omega_j)=\frac{1}{\sqrt{N}}\sum_i u^{\alpha}(t_i) \exp(-\mathrm{i} \omega_j t_i)
\end{eqnarray}
with $\omega_j=\frac{2\pi j}{\Delta t    N}$.

To obtain a discrete time trace $\beta(t)$ corresponding to the spectrum density $S(\omega)$, we take the inverse Fourier transform of the product of $\tilde{u}^{\alpha}(\omega_j)$ and $\sqrt{S'(\omega_j) }$:
\begin{eqnarray}\label{eq:beta}
\beta ^\alpha_l=\beta ^\alpha(t_l)= \sum_{j=1}^N \frac{1}{\sqrt{N}} \tilde{u}^{\alpha}(\omega_j) \sqrt{S'(\omega_j)} \exp(\mathrm{i} \omega_j t_l), \label{eq:generationBeta}
\end{eqnarray}
 where $t_l=l \Delta t$ and $S'(\omega_j)$ corresponds to $S(\omega_j)/\Delta t$ folded about $\omega_j=N/2+1$:
\begin{eqnarray}
S'(\omega_j)=\begin{cases}
	S(\omega_j)/\Delta t & j=1,2 ... N/2+1\\
	S(\omega_{N+2-j})/\Delta t & j=N/2+2 ... N\\
	\end{cases}.
	\label{newnoiseSpec} 
\end{eqnarray}

\section{Method of the Numerical simulation}
The two-level Hamiltonian describing a single spin under microwave excitation $\omega_1 \sin (\omega_0 t + \phi)$, setting $/hbar=1$, can be written in the rotating frame of its Larmor frequency $\omega_0$ as
\begin{eqnarray}
H_\mathrm{R}(t)	&=& \omega_1(t)\left(\frac{\sigma_x}{2}\cos \phi(t) -\frac{\sigma_y}{2} \sin \phi(t) \right) +\eta(t) \frac{\sigma_z}{2} \label{HamiltonianRotatingFrameMoreGerarality} 
\end{eqnarray}
with $\eta(t)=\gamma_e \delta b(t)$ and $\delta b(t)$ a fluctuating magnetic field along the same direction as the quantization axis, $\hat{z}$. We take $\omega_1$ and $\phi$ to be a function of time since we turn on and turn off the microwave excitation and change the microwave phase according to the quantum gates.

Here we treat the evolution of the density operator according to Eq.~\eqref{HamiltonianRotatingFrameMoreGerarality} numerically, assuming that $H_\mathrm{R}(t)$ can be considered as constant for a short time $\Delta t$:
\begin{eqnarray}
H_\mathrm{R}(t_i)	&=& \omega_{1i}\left(\frac{\sigma_x}{2}\cos \phi_i -\frac{\sigma_y}{2} \sin \phi_i \right) +\eta_i \frac{\sigma_z}{2}, \label{HamiltonianRotatingFrameMoreGeraralityDiscrete} 
\end{eqnarray}
with $t_i=i\Delta t$.
Then the density operator at time $t=n\Delta t$ becomes 
\begin{eqnarray}
\rho(n\Delta t)=\left\langle \bigcirc_{i=1}^n \mathcal{U}_i (\rho_0) \right\rangle_\alpha= \frac{1}{M} \sum_\alpha^M \left[\bigcirc_{i=1}^n \mathcal{U}_i^\alpha (\rho_0)  \right],  \label{TimeEvolutionDiscreteRho} 
\end{eqnarray}
where $\rho_0$ is the initial state and $\mathcal{U}_i$ is the superoperator representation:
\begin{eqnarray}
\mathcal{U}_i (\rho) =U_i^\dag \rho U_i  \label{superoperator} 
\end{eqnarray}
\begin{eqnarray}
\bigcirc_{i=1}^n \mathcal{U}_i (\rho)=\mathcal{U}_n \circ ... \circ \mathcal{U}_3 \circ \mathcal{U}_2 \circ \mathcal{U}_1 (\rho) =U_n^\dag...U_3^\dag U_2^\dag U_1^\dag \rho U_1 U_2 U_3...U_n  \label{SuperoperatorSum} 
\end{eqnarray}
with 
\begin{eqnarray}
U_i=\exp \left(-\mathrm{i} H_\mathrm{R}(t_i)    \Delta t \right).
\end{eqnarray}
The $\omega_{1i}$ is a step function with the value of the Rabi frequency $\omega_1$: it is set to $\omega_1$ when microwaves are applied and 0 otherwise.
%
$\phi_i$ is set depending on the phase of the applied microwave. $\eta_i$ is the sum of the static noise $\beta_0^\alpha$ and the high-frequency noise $\beta_i^\alpha$:
\begin{eqnarray}
\eta_i^\alpha=\beta_0^\alpha+\beta_i^\alpha,  \label{betacombi} 
\end{eqnarray}
where $\beta_0^\alpha$ has a Gaussian distribution with standard deviation 0.268 MHz (FWHM = 0.63 MHz) and $\beta_i^\alpha$ is generated by Eq.~\eqref{eq:generationBeta} with $S(\omega)=\frac{A}{1+(\omega \tau_c)^2}$, $A=2.5 \cdot 10^4 \mathrm{rad}^2$s$^{-1}$ and $\tau_c$=2.46 $\mu$s as discussed in the main text. In Eq.~\ref{betacombi}, we explicitly write that $\eta$ gives a different value with different $\alpha$. The simulations for randomized benchmarking and dynamical decoupling were performed with $M=1000$, $\Delta t=5$ ns, $N=200000$, $\omega_1=\frac{2\pi}{360 \mathrm{ns}}$ unless otherwise stated. From $\Delta t=5$ ns, the highest frequency taken into account for $\beta_i^\alpha$ is 200 MHz and from $N\cdot \Delta t$=1 ms, the lowest frequency is 1 kHz. The frequency range of 1 kHz-200 MHz is a larger range than the one explored with the dynamical decoupling experimentally.


\section{Numerical simulation for Randomized Benchmarking}
The main results of the simulation are shown in the main text. 

Here the average gate fidelity for a single gate ($m=1$) with the same noise used for Fig.~4(b) is calculated. The initial quantum state $\rho_0$, after undergoing a depolarizing channel produced by applying one set of imperfect Clifford gates and perfect inverse Cliffird gates, becomes 
\begin{eqnarray}
\rho=\frac{1}{24}\sum_{i=1}^{24}{{\mathcal{C}_i^{\mathrm{ideal}}}^\dag \circ \mathcal{C}_i^\mathrm{real}} ( \rho_0)=p\rho_0+(1-p) \frac{\1}{2}, \label{eq:twirl_depolarization2}
\end{eqnarray}
where $\mathcal{C}_i^\mathrm{ideal}$ is the superoperator of a perfect Clifford gate without noise and $\mathcal{C}_i^\mathrm{real}$ is the superoperator of an imperfect Clifford gate affected by the noise described by Eq.~\eqref{betacombi}.
By taking the POVM operator as $E=\Ketbra{\uparrow}$ and $\rho_0=\Ketbra{\downarrow}$ \footnote{In fact, not depending on how $E$ and $\rho_0$ are taken, the calculated gate fidelity should be the same.}, the measured spin-up probability becomes 
\begin{eqnarray}
P_{\uparrow}&=&\mathrm{Tr} \left[ E \rho \right]=\mathrm{Tr} \left[ \Ketbra{\uparrow} \left(p\Ketbra{\downarrow}+(1-p) \frac{\1}{2} \right)\right]=\frac{1-p}{2}. \label{eq:PupCliffm1gate}
\end{eqnarray}
Finally the average gate fidelity for an isolated gate ($m=1$) is calculated as $1-P_{\uparrow}/1.875=97.8$ \% (There are 45 single qubit gates used across 24 Cliffords \cite{Barends2014}. In order to obtain the average gate fidelity of a single qubit gate, the error is divided by 45/24.), which is lower than the average gate fidelity per single gate $\approx$ 99 $\%$ obtained by the fits to the curves in Fig.~4(a,b), which is the averaged gate fidelity when several gates are concatenated ($m=2$ to $10$). The improvement of the gate fidelity with a higher number of gates stems from the fact that some RB sequences have the effect of partial error suppression of low-frequency noise \cite{Epstein2014,Ball2015}. We consider the averaged gate fidelity over the gates for $m=2$ to $10$ as the fidelity of interest because the average gate fidelity per single gate for $m=1$ cannot be measured by randomized benchmarking and since more than 2 gates are usually applied in the real quantum computation.

\begin{figure}[h!]
\centering
\includegraphics[width=9cm]  {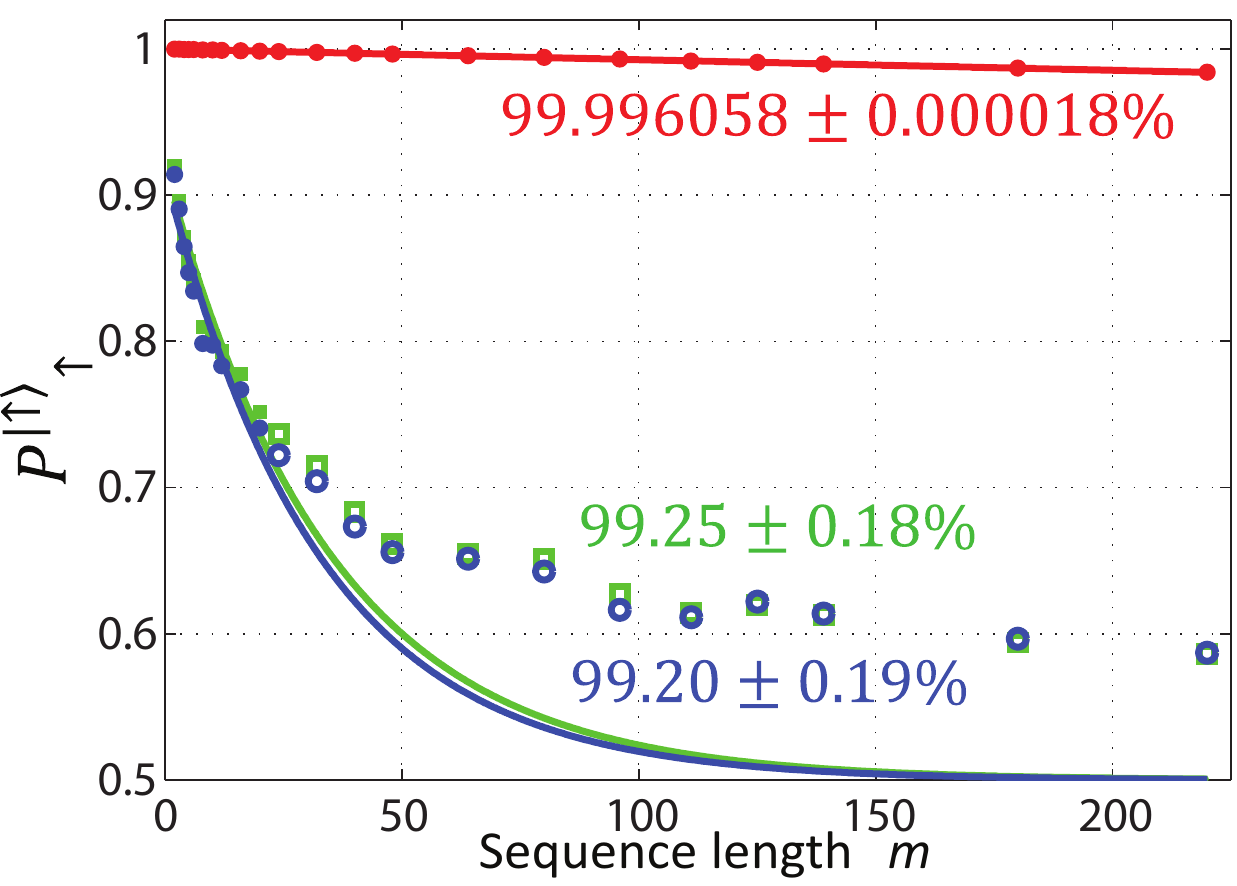}
\caption{\label{fig:RBsimu} Numerical simulation of randomized benchmarking on $P_\uparrow^{\left|\uparrow \right\rangle}$ for 3 cases: considering only the high-frequency noise (red circles), considering only the (quasi-)static noise (green squares) and considering both (blue circles). The solid lines are fits to the simulation results using the form $ap^m+0.5$,  with $t_p<$8 $\mu$s for the later two cases (filled red and blue circles). The extracted average gate fidelity per single gate are shown using the same colors as data points.}
\end{figure}

 Fig.~\ref{fig:RBsimu} shows the results of the numerical simulations of the spin-up probability with spin-up as the target state $P_\uparrow^{\left|\uparrow \right\rangle}$ including only the high-frequency noise $\delta b'(t)$ (red circles), only the (quasi-)static noise $\delta b_0$ (green squares), and both $\delta b_0+\delta b'(t)$ (blue circles). With only the high-frequency noise, the decay is very slow. In this case, the flat region of the Lorentian spectrum mainly contributes to the gate errors and thus the noise can be regarded as uncorrelated, which results in the whole decay curve following a single power law and the extracted average gate fidelity per single gate from the fit of $ap^m+0.5$ is $>99.99\%$, which means that the high-frequency noise $\delta b'(t)$ would not limit the gate fidelity with the isotopically purified Si sample. The decay curve with only the (quasi-)static noise and that with the combination of two show almost the same decay. It indicates that the (quasi-)static noise is mainly responsible for the simulated and measured gate errors.

\section{Numerical simulation for Dynamical Decoupling}
\begin{figure}[h!]
\includegraphics[width=17cm] {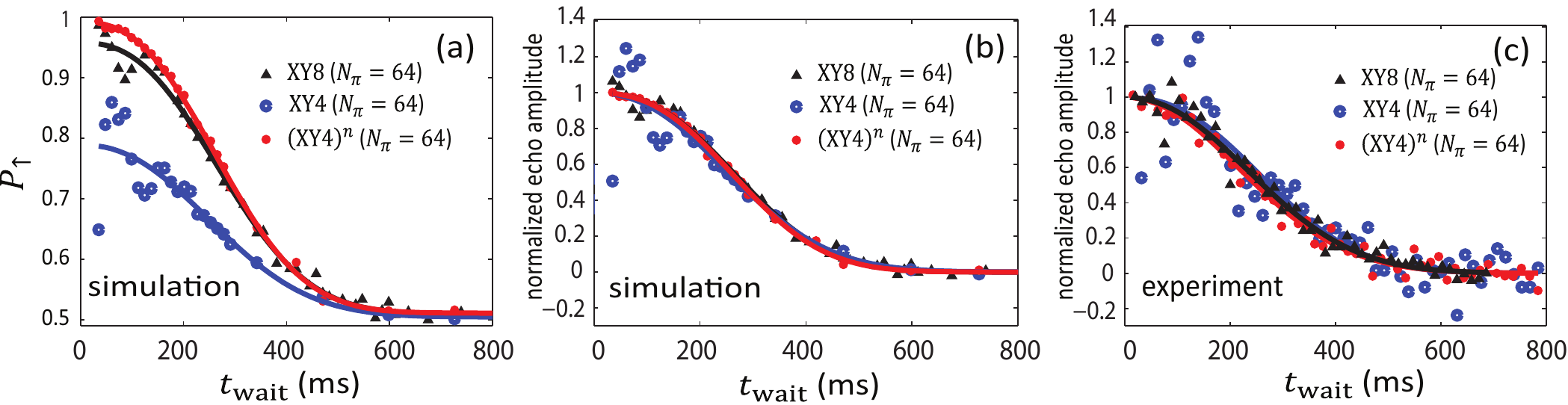}
\caption{\label{fig:DDsimuN64} (a-c) the echo decays with XY4, XY8 and (XY4)$^n$ for $N_\pi=64$ (a) echo decays calculated from the numerical simulation. The solid lines are the fits with Eq.~(4). $T_2= 320\pm 92$, $325 \pm 12$ and $319 \pm 7$ $\mu$s and $\alpha=2.28 \pm 1.78$, $2.50 \pm 0.30$ and $2.53 \pm 0.18$ are obtained, for XY4, XY8 and (XY4)$^n$, respectively. (b) The echo decay data are normalized according to the fits with Eq.~(4). (c) Measured echo decay normalized according to the fits with Eq.~(4).}
\end{figure}

We have used 3 different types of dynamical decoupling pulse sequences: XY4\cite{Maudsley1986}, XY8\cite{Gullion1990} and (XY4)$^n$\cite{Alvarez2012}. CPMG timing is used as the time interval between pulses for all the three sequences and so their filter functions are the same, while the phases of the $\pi$ pulses, shown in Eq.~(\ref{XY4}-\ref{XY4n}) for $N_\pi=64$, are different.
\begin{eqnarray}
\mathrm{XY4}&=&\: XYXY	\:	 XYXY	\:	 XYXY \:		XYXY \: 
						  	\:XYXY	\:	XYXY	\:	 XYXY	 \:		XYXY 	\nonumber\\
&& XYXY 	\:	 XYXY	\:	 XYXY \:		XYXY \:  
						  	\: XYXY	\:	XYXY 	\:	XYXY	 \:		XYXY \: \label{XY4} \\
\mathrm{XY8}&=&\: XYXY	\:	 YXYX	\:	 XYXY \:		YXYX \: 
						  	\:XYXY	\:	YXYX	\:	 XYXY	 \:		YXYX 	\nonumber\\
&& XYXY 	\:	 YXYX	\:	 XYXY \:		YXYX \:  
						  	\: XYXY	\:	YXYX 	\:	XYXY	 \:		YXYX \:   	\label{XY8} \\
\mathrm{(XY4)}^3&=&\: XYXY	\:	 X \Ybar X \Ybar 	\:	 \Xbar \Ybar \Xbar \Ybar	 \:		\Xbar Y \Xbar Y \: 
						  	\: X \Ybar X \Ybar 	\:	 X Y X Y 	\:	 \Xbar Y \Xbar Y	 \:		\Xbar \Ybar \Xbar \Ybar 	\nonumber\\
&& \Xbar \Ybar \Xbar \Ybar 	\:	 \Xbar Y \Xbar Y 	\:	 X Y X Y	 \:		X \Ybar X \Ybar \:  
						  	\: \Xbar Y\Xbar Y	\:	 \Xbar  \Ybar \Xbar  \Ybar 	\:	 X \Ybar X \Ybar	 \:		X Y X Y \:  \label{XY4n}
\end{eqnarray}
(XY4)$^n$ is built from the concatenation of XY4 pulse sequences. $n$ stands for the concatenation level and $n=3$ for $N_\pi=64$. (XY4)$^n$ is sometimes called virtual-CDD (vCDD). (XY4)$^1$ is equal to the XY4 pulse sequence. (XY4)$^n$ is made by concatenating (XY4)$^{n-1}$ 4 times with $X$ or $Y$ pulses inserted between (XY4)$^{n-1}$. The inserted $\pi$ pulses are replaced by virtual pulses by toggling the frame of Hamiltonian. This method is more robust against pulse imperfections  \cite{Alvarez2012}. Increasing $N_\pi$, the pulse errors are accumulated and so its effect becomes bigger. Both our experiment and the simulation show that (XY4)$^n$ works better in terms of suppressing the effect of the pulse imperfections than other pulse sequences (Fig.~\ref{fig:DDsimuN64}). As discussed in the section on randomized benchmarking, the quasi-static noise is the dominant source for the gate errors. Such errors can be partly canceled out by applying an appropriate combination of pulse sequences like (XY4)$^n$.

 Fig.~\ref{fig:DDsimuN64}(a-c) show simulated echo decays with XY4, XY8 and (XY4)$^n$ for $N_\pi=64$. As seen in Fig.~\ref{fig:DDsimuN64}(a), the visibility for (XY4)$^n$ is much better than for XY4 and also slightly better than for XY8. With XY4 and XY8, the decay curves exhibit oscillations in the beginning in both simulation and experiment, while this does not happen for (XY4)$^n$. Fig.~\ref{fig:DDsimuN64}(b) and (c) show the normalized decay curves at $t_\mathrm{wait}=0$ from the fits for the simulation and the experiment. The fitted lines to the decay curves for  XY4, XY8 and (XY4)$^n$ match with each other very well and the $T_2$ time (see caption in Fig.~\ref{fig:DDsimuN64}) is almost equal for all the sequences. This is expected given that the filter function is the same. However the standard errors on the fitting parameters are more than 10 times higher with XY4 than with (XY4)$^n$, due to the reduced visibility and to the oscillations for short $t_\mathrm{wait}$. In conclusion, even if it is still possible to determine $T_2$ in the presence of pulse imperfections, one needs to use a pulse sequence which is robust to the pulse imperfections like (XY4)$^n$ in order to determine $T_2$ with high accuracy. In addition, in terms of quantum information processing, it is important to preserve a quantum state. (XY4)$^n$ is the preferred pulse sequence in this sense since it succeeds in filtering out the high-frequency noise while canceling out the pulse errors due to the low-frequency noise at the same time.

\begin{figure}[h!]
\centering
\includegraphics[width=\textwidth] {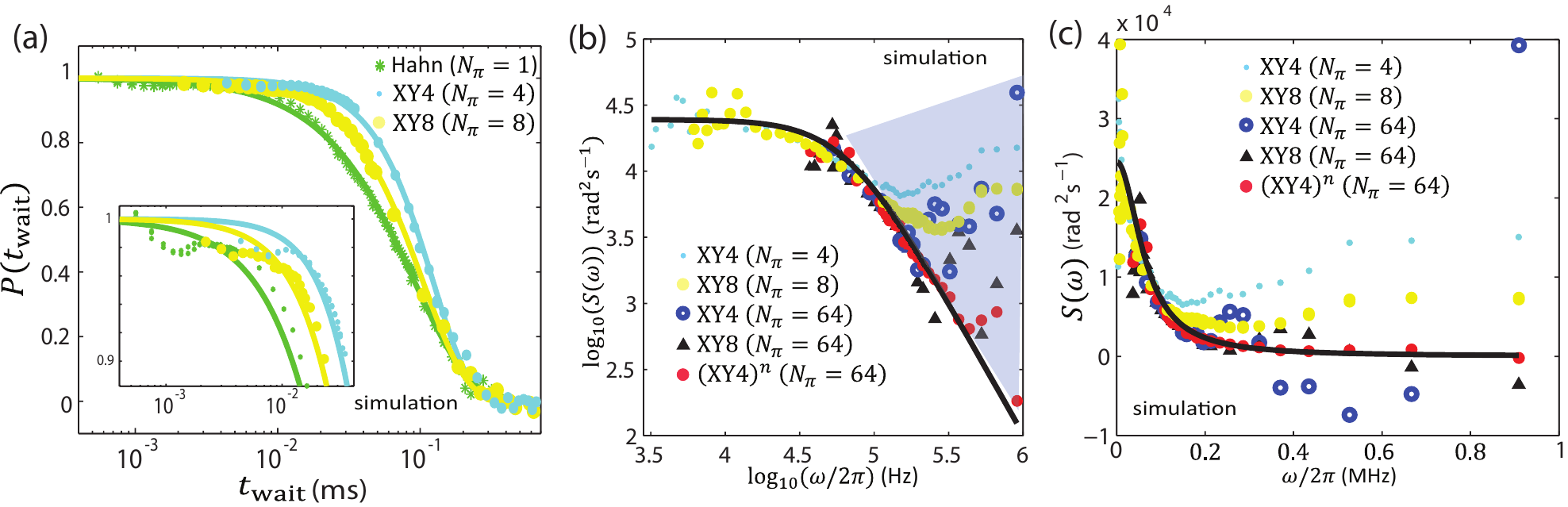}
\caption{\label{fig:DDsimuNvarious}(a) Simulated echo decays with Hahn ($N_\pi=1$), XY4 ($N_\pi=4$) and XY8 ($N_\pi=8$). The inset shows the zoom-up of the beginning of the decays. The solid lines present the fits with Eq.~(4). (b,c) Noise spectrum extracted from numerical simulated echo decays using XY4 ($N_\pi=4$), XY8 ($N_\pi=8$), XY4 ($N_\pi=64$), XY8 ($N_\pi=64$) and (XY4)$^n$ ($N_\pi=64$) pulse sequences. The solid black line is the noise spectrum which is used in the simulation to produce the echo decays. In (b), decimal logarithmic scales are used along x and y axis.}
\end{figure}

In Fig.~\ref{fig:DDsimuNvarious}(a), the simulated echo decays for Hahn ($N_\pi=1$), XY4 ($N_\pi=4$) and XY8 ($N_\pi=8$) are shown. Fig.~\ref{fig:DDsimuNvarious}(b,c) show the noise spectrum extracted from the simulated echo decay curves presented in Fig.~\ref{fig:DDsimuN64}(b) and Fig.~\ref{fig:DDsimuNvarious}(a) using the same methods as used in Fig.~3(d).

The solid black lines in Figs.~S8(b) and (c) represents the noise spectrum used to produce the echo decays in the numerical simulations in Fig.~S8(a). The deviation of the extracted noise spectrum from the input noise spectrum is due to the oscillations seen in the echo decays and normalization problems. The noise spectrum extracted form (XY4)$^n$ echo decay follows the input noise spectrum the best. In the region of $10^5<\omega/2\pi$, the deviations of the extracted noise for XY4 ($N_\pi=4$) and XY8 ($N_\pi=8$) are above the input noise spectrum (Fig.~\ref{fig:DDsimuNvarious}(b,c)). This is related to normalization problems. The inset of Fig.~\ref{fig:DDsimuNvarious}(a) shows that the data points for each echo decay are below each respective fitted line for short waiting times. This results in an extracted noise spectrum that is higher than the input noise spectrum, in particular for the high frequency range (as seen in the shaded region in Fig.~\ref{fig:DDsimuNvarious}(b)). This problem arises from the difficulty of proper normalization of the simulated data points and for the artifact of the oscillations at short $t_\mathrm{wait}$. Thus even if at higher frequencies, Eq.~[3] with $\alpha \sim 2$ seems to work better to capture the power spectrum in Fig.~3(d) of the main text, it happens due to this normalization problem. Besides that, there is another normalization problem. When the echo decays are normalized, some data points where $t_\mathrm{wait}$ is short (for example, 4 points for XY4 ($N_\pi=64$) and 2 points for XY8 ($N_\pi=64$),) of Fig.~\ref{fig:DDsimuN64}(b) go above 1. In these cases, for the logarithmic plot in Fig.~\ref{fig:DDsimuNvarious}(b), $-\mathrm{log}P(t_\mathrm{wait})$ in Eq.~\eqref{SApprox} is replaced with $|-\mathrm{log}P(t_\mathrm{wait})|$. These points are not reliable. 

\section{additional data for dynamical decoupling}
\begin{figure}[h!]
\centering
\includegraphics[width=15cm]  {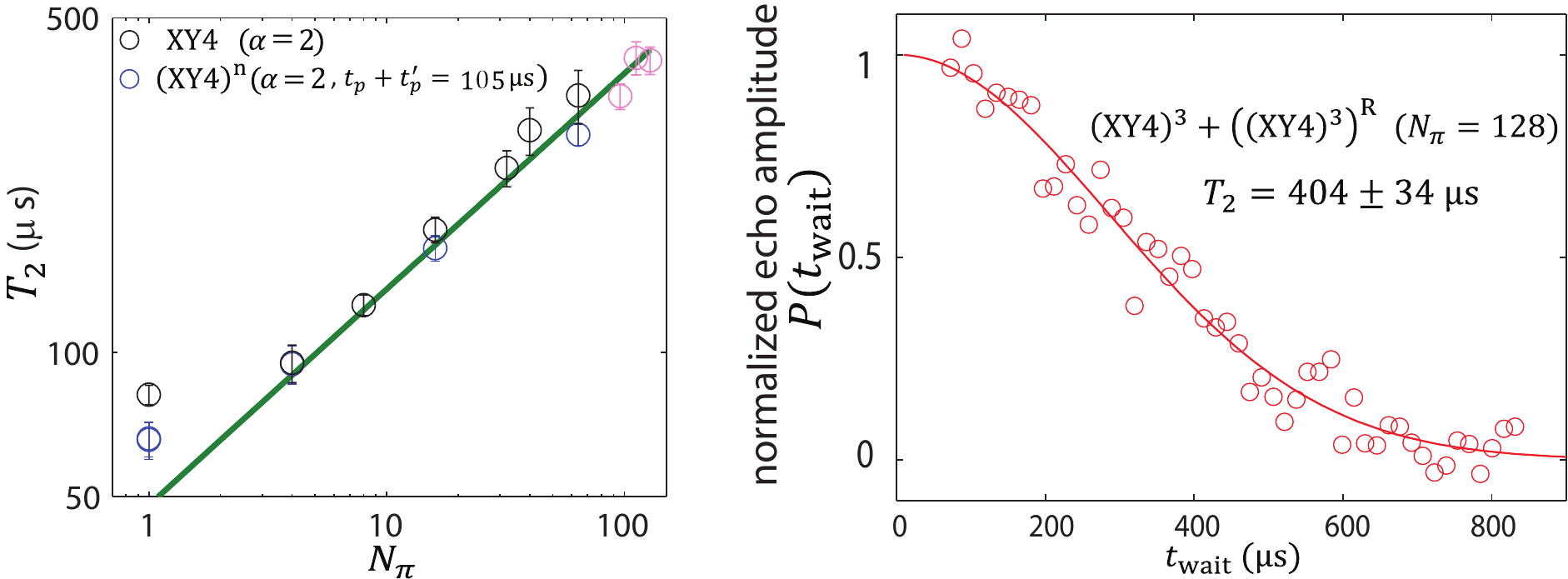}
\caption{\label{fig:T2vsN_and_N128decay}(a) Coherence time, $T_2$, as a function of the number of $\pi$ pulses using XY8 without (black circles) and with (blue circles) off-resonance microwave. The green line is the same as in Fig.~3(c). (b) Echo decay with 128 $\pi$ pulses. The fit with a Gaussian curve (red line) gives $T_2 \approx$400 $\mu$s.}
\end{figure}

Here we provide additional dynamical decoupling data. First, the blue circles in Fig.~\ref{fig:T2vsN_and_N128decay} show data analogous to purple circles in Fig.~3(c) but with an off-resonance microwave burst added in order to keep the total microwave burst time fixed. The $T_2$ does not significantly change when adding the off-resonance microwave. Second, we show as black circles the echo decay using the XY4 pulse sequence. The green line and pink circles are the same as in Fig.~3(c). The pink circles show $T_2$ with the number of $\pi$ pulses, 96, 112 and 128 for $\alpha=2$. In Fig.~\ref{fig:T2vsN_and_N128decay}(b), the red circles shows the decay curve with $N_\pi=128$ and the red line shows the fit with Eq.~(4) for $\alpha=2$. From this fit, $T_2=404 \pm 34$ $\mu$s is obtained. 
 $(\bullet)^R$ stands for applying the pulse sequence $(\bullet)$ in reverse order. We note that extending the (XY4)$^n$ pulse sequence beyond $N_\pi=64$ would take us immediately to 256 pulses. The pulse sequences used for $N_\pi=96$ is (XY4)$^2$+(XY4)$^3$+((XY4)$^2$)$^R$, for $N_\pi=112$ is repeating (XY4)$^2$ 7 times, and for $N_\pi=128$ is (XY4)$^3$+((XY4)$^3$)$^R$.

\bibliography{C:/Users/tud205337/Dropbox/Research/Tex/MyCollection}

\end{document}